\documentclass[twocolumn]{aastex631}

\usepackage{cleveref}
\usepackage{booktabs}
\usepackage{longtable}
\usepackage{array}

\makeatletter
\usepackage{etoolbox}
\patchcmd\H@refstepcounter{\protected@edef}{\protected@xdef}{}{}
\makeatother

\newcolumntype{i}{>{\scriptsize}r}

\shorttitle{Compact Symmetric Objects}
\shortauthors{Kiehlmann et al.}

\graphicspath{{./}{figs/}}

\crefname{equation}{Eq.}{Eqs.}
\Crefname{equation}{Equation}{Equations}
\crefname{figure}{Fig.}{Figs.}
\Crefname{figure}{Figure}{Figures}
\crefname{table}{Table}{Tables}
\Crefname{table}{Table}{Tables}
\crefname{section}{Section}{Sections}
\Crefname{section}{Section}{Sections}

\usepackage[nolist,nohyperlinks]{acronym}
\begin{acronym}
    \acro{ads}[ADS]{Astrophysics Data System}
    \acro{agn}[AGN]{Active Galactic Nucleus}
    \acroplural{agn}[AGN]{Active Galactic Nuclei}
    \acro{bologna}[Bologna]{Bologna Complete Sample of Nearby Radio Sources}
    \acro{cd}[CD]{Compact Double}
    \acro{cj1}[CJ1]{First Caltech Jodrell Bank}
    \acro{cjf}[CJF]{Caltech Jodrell Bank flat-spectrum}
    \acro{class}[CLASS]{Cosmic Lens All-Sky Survey}
    \acro{cso}[CSO]{Compact Symmetric Object}
    \acro{css}[CSS]{Compact Steep Spectrum}
    \acro{ddrg}[DDRG]{Double-Double Radio Galaxy}
    \acroplural{ddrg}[DDRGs]{Double-Double Radio Galaxies}
    \acro{emerlin}[eMERLIN]{Enhanced Multi Element Remotely Linked Interferometer Network}
    \acro{fr}[FR]{Fanaroff–Riley}
    \acro{fsrq}[FSRQ]{Flat Spectrum Radio Quasar}
    \acro{fwhm}[FWHM]{Full Width at Half Maximum}
    \acro{ps}[PS]{Peaked Spectrum}
    \acro{hfp}[HFP]{High-Frequency Peaker}
    \acro{lso}[LSO]{Large Symmetric Object}
    \acro{mso}[MSO]{Medium Symmetric Object}
    \acro{ned}[NED]{NASA/IPAC Extragalactic Database}
    \acro{ovro}[OVRO]{Owens Valley Radio Observatory}
    \acro{pah}[PAH]{Polycyclic Aromatic Hydrocarbon}
    \acro{pr}[PR]{Pearson Readhead}
    \acro{ps}[PS]{Peaked-Spectrum}
    \acro{prcj1}[PR+CJ1]{Pearson Readhead + First Caltech Jodrell Bank}
    \acro{pw}[PW]{Peacock Wall}
    \acro{rfc}[RFC]{Radio Fundamental Catalog}
    \acro{vips}[VIPS]{VLBA Imaging and Polarimetry Survey}
    \acro{vla}[VLA]{Very Large Array}
    \acro{vlba}[VLBA]{Very Long Baseline Array}
    \acro{vlbi}[VLBI]{Very Long Baseline Interferometry}
\end{acronym}

\begin{document}

\title{Compact Symmetric Objects - I\\
Towards a Comprehensive Bona Fide Catalog}

\correspondingauthor{Anthony Readhead}
\email{acr@caltech.edu}

\author{S. Kiehlmann}
\affiliation{Institute of Astrophysics, Foundation for Research and Technology-Hellas, GR-70013 Heraklion,Greece}
\affiliation{Department of Physics and Institute of Theoretical and Computational Physics, University of Crete, 70013 Heraklion, Greece}
\author{M. L. Lister}
\affiliation{Department of Physics and Astronomy, Purdue University, 525 Northwestern Avenue, West Lafayette, IN 47907, USA}
\author{A. C. S Readhead}
\affiliation{Owens Valley Radio Observatory, California Institute of Technology,  Pasadena, CA 91125, USA}
\author{I. Liodakis}
\affiliation{Finnish Center for Astronomy with ESO, University of Turku, Vesilinnantie 5, FI-20014, Finland}
\author{S. O'Neill}
\affiliation{Owens Valley Radio Observatory, California Institute of Technology,  Pasadena, CA 91125, USA}
\author{T. J. Pearson}
\affiliation{Owens Valley Radio Observatory, California Institute of Technology,  Pasadena, CA 91125, USA}
\author{E. Sheldahl}
\affiliation{Department of Physics and Astronomy, University of New Mexico, Albuquerque, NM 87131, USA}
\author{A. Siemiginowska}
\affiliation{Center for Astrophysics|Harvard and Smithsonian, 60 Garden St., Cambridge, MA 02138, USA}
\author{K. Tassis}
\affiliation{Institute of Astrophysics, Foundation for Research and Technology-Hellas, GR-70013 Heraklion,Greece}
\affiliation{Department of Physics and Institute of Theoretical and Computational Physics, University of Crete, 70013 Heraklion, Greece}
\author{G. B. Taylor}
\affiliation{Department of Physics and Astronomy, University of New Mexico, Albuquerque, NM 87131, USA}
\author{P. N. Wilkinson}
\affiliation{Jodrell Bank Centre for Astrophysics, University of Manchester, Oxford Road, Manchester M13 9PL, UK} 

\begin{abstract}
   \acp{cso} are jetted \acp{agn} with overall projected size $<1\,\mathrm{kpc}$. The classification was introduced to distinguish these objects from the majority of compact jetted-\acp{agn}  in centimeter wavelength very long baseline interferometry observations, where the observed emission is relativistically boosted towards the observer. The original classification criteria for \acp{cso}  were: (i) evidence of emission on both sides of the center of activity, and (ii)  overall size $<1\,\mathrm{kpc}$. However some relativistically boosted objects with jet axes close to the line of sight appear symmetric and have been mis-classified as \acp{cso}, thereby undermining the \ac{cso} classification. This is because two essential  \ac{cso} properties, pointed out in the original papers, have been neglected:  (iii) low  variability, and (iv) low apparent speeds along the jets. As a first step towards creating a comprehensive catalog of ``bona fide'' \acp{cso}, we identify 79~bona fide \acp{cso},  including 15 objects claimed as confirmed CSOs here for the first time, that match  the \ac{cso} selection criteria.
    This sample of bona fide \acp{cso} can be used for astrophysical studies of \acp{cso} without contamination by mis-classified  \acp{cso}. We show that  the fraction of \acp{cso} in  complete flux density limited AGN samples with  S$_{\rm 5\,GHz}>700\,\mathrm{mJy}$ is between  $(6.8\pm1.6)$\% and $(8.5\pm1.8)$\%.

\end{abstract}


%------------------------------------------------------------------------------
\section{Introduction}
\label{sec:intro}

% series of five papers:

% literature background:

The discovery that most compact ($<1\,\mathrm{kpc}$) radio sources in cm wavelength surveys have asymmetric, one-sided jet morphology, with a bright compact core at one end of a steep spectrum jet \citep{1977Natur.269..764W,1978Natur.276..768R,1980IAUS...92..165R}, provided powerful support for the hypothesis that the observed radio emission regions,  especially  in flat spectrum radio sources, are strongly boosted by relativistic beaming \citep{1966Natur.211..468R,1967MNRAS.135..345R}. This quickly became a crucial pillar of our understanding of jetted-active galactic nuclei (jetted-\acp{agn}) \citep{1984RvMP...56..255B,2019ApJ...874...43L,2019ARAA..57..467B}.

It was a surprise, therefore, when, in 1980, Phillips \& Mutel discovered  a  class of compact  extragalactic radio sources that, unlike the majority of jetted-\acp{agn}, do not have an apparent asymmetric, one-sided jet morphology \citep{1980ApJ...236...89P, 1980ApJ...241L..73M}, but are \acp{cd}. Phillips \& Mutel suggested that \acp{cd} are young radio sources associated with \acp{agn}. 

The \ac{agn} B3 0713+439 provided the first clear evidence of symmetry in these objects \citep{1984IAUS..110..131R}, since there are clearly two components on opposite sides of the nucleus.   The first \ac{vlbi} survey of a complete sample\footnote{A ``complete sample'' is defined to be a sample that includes all objects down to a given flux density limit over a given area of sky \citep{1968MNRAS.139..515P,1968ApJ...151..393S,1970MNRAS.151...45L}} of extragalactic radio sources was that of the 65~\acp{agn} studied by \citet{1988ApJ...328..114P},  hereafter PR88, who found three \acp{cd} with steep spectra and concluded  that they ``seem to be a completely different kind of object. The evidence suggests that the radio structures that we see are not parts of cores or jets''. It was clear from early on, therefore, that there was something very different about these Compact {\it Symmetric\/} Objects (CSOs). The fraction of \acp{cso}, $\sim 10$\% in the \ac{pr} complete sample, and the sizes, $\sim 100\,\mathrm{pc}$, imply that they are either very slowly expanding old objects, or very short-lived objects. 

The original motivation for the \ac{cso} classification is given by  \cite{1994ApJ...432L..87W}, hereafter W94,  their properties are summarized  succinctly  by \citet{1993AAS...182.5307R}, hereafter R93, 
  and their relationship to other larger radio sources is discussed in \citet{1994cers.conf...17R}, hereafter R94, while a detailed discussion of \acp{cso}  as a class is given in \citet{1996ApJ...460..612R}, hereafter R96 . In defining \acp{cso}, W94 and R96 were aiming to avoid contamination  by objects in which the observed emission is strongly beamed towards the observer -- see the third property, after morphology and size, listed by R93 --  by imposing a requirement of symmetry about the center of activity. However,   understandable but unfortunate, misidentifications of \acp{agn} as \acp{cso} have eroded the \ac{cso} class to the point where the phenomenology of the \ac{cso} class has been obscured. Before undertaking this review,  we considered carefully whether it is worth resurrecting the \ac{cso} class, or whether its value has diminished with time since it was first proposed, and we concluded that if the original intent of the \ac{cso} classification could be recaptured, then it would be well worth doing. As pointed out by \citet{2016MNRAS.459..820T}, the \ac{cso} category is a much more physically motivated classification  than the  Peaked Spectrum (PS)\footnote{In this paper we follow the lead of  \citet{2021AARv..29....3O} in their comprehensive review of peaked spectrum sources, and refer to Gigahertz Peaked Spectrum (GPS) sources as Peaked Spectrum (PS) sources} and \ac{css} classifications,  the former of which are highly contaminated with  objects in which the observed radiation is strongly affected by relativistic beaming \citep{2009AN....330..128T}, and the latter of which include a diverse variety of AGN morphological types.  

We began this program with the sole intention of re-defining the \ac{cso} classification criteria, and beginning the task of identifying the ``bona fide'' \acp{cso} in the literature. However, as we progressed, we found compelling evidence that  most  \acp{cso} form a distinct class of jetted-\acp{agn} that are different in their nature and origin from all other classes of jetted-\acp{agn}. This fact had already been pointed out  by PR88, R94, and R96, but it has never taken root. We trust that the new evidence we present in this study will remedy that situation.  We also found that,   while most of the \acp{cso} observed thus far are short-lived compared to the classes of larger radio sources, they are certainly not all young, and that  \acp{cso} can  reliably be classified into four distinct morphological classes, of which three   most likely represent the ``early life'', ``mid-life'' and ``late-life'' phases of  an evolutionary trajectory. The evidence suggests  that most high-luminosity CSOs might originate in a  single fuelling event  through the capture of a single star by an SMBH, as first suggested by R94, and again more recently by \citet{2012ApJ...760...77A}.

All of these findings amount to a significant change in our  understanding of jetted-\acp{agn}, and provide a new avenue of attack on relativistic jets, which has the potential to probe the physics of the generation of the jets and their relationship to the accretion disk and the surrounding medium in the centers of activity in \acp{agn} in an entirely new way.  

Our findings are reported in a set of three papers.  In this, the first paper, hereafter Paper~1 in this series, we describe a large and intensive CSO literature search that we have carried out, and the selection of 79~``bona fide'' \acp{cso}. In the second paper (Kiehlmann et al., in press), hereafter Paper~2 in this series, we present what we consider to be compelling evidence that \acp{cso} form a distinct class of jetted-\acp{agn}.  In the third paper (Readhead et al. in press), hereafter Paper~3 in this series, we introduce a 4-way sub-classification of \acp{cso} and discuss the origin and evolution of  the majority of  \acp{cso}, and we show that  most CSOs might  well be fuelled by the capture of single stars by  dormant spinning supermassive black holes in a the nuclei of elliptical galaxies.

\citet{2016MNRAS.459..820T} carried out the most extensive survey of \acp{cso} to date. They found  two classes of \acp{cso} that are analogous to the \ac{fr} classes \citep{1974MNRAS.167P..31F} of the large double radio sources: in one class of \ac{cso} the leading (or outermost) edges of the sources are fainter than the central parts of the source, i.e., they have ``edge-dimmed'' morphology,  and this class also has low luminosity, which we designate as \ac{cso}~1 objects. The other class has edge-brightened morphology, which we designate as \ac{cso}~2 objects. We find strong confirming evidence of these two distinct classes of \acp{cso}, so we have classified our sample accordingly.  In \cref{fig:morphology} we show three typical examples of the morphologies of these CSO 1 and CSO 2 classes. In Paper~3 we show that  the vast majority of \ac{cso}~2s can be subdivided into three sub-classes that follow an evolutionary path.   We base our  classification of CSOs into four sub-classes on morphology alone.

It is important at the outset to mention the limitations of the present study. Clearly, the most powerful and efficient way to study CSOs is to carry out a large, unbiased VLBI survey covering many thousands of jetted-AGN that will produce a complete sample of CSOs  significantly larger than that available at present.  We are engaged in just such a study, but it will take several years to complete.

There is, however, no need to wait for the outcome of this survey to make progress on the study of CSOs, because many thousands of hours of VLBI observing time have been devoted to the study of CSOs, and much can be gained from using the invaluable results  in the existing literature.  However, it should be borne in mind that a large fraction of the CSO candidates in the literature come from a wide variety of observing programs, many of  which have significant selection effects.  For these reasons, it is not possible to use all of the  bona fide CSOs identified in this study for all of the purposes that we address in these three papers.  The lack of spectroscopic redshifts for $\sim$30\% of CSOs is one obvious example.  Other obvious examples are the different observing frequencies that have been used, the different flux density limits of different studies, and the incompleteness of variability surveys.

It should be clear that it is possible to make significant progress in the study of CSOs only due to the enormous effort that has been put into the study of these objects by various groups. Since this is not a review, we do not attempt a description of this earlier work, but our extensive bibliography attests to it.  Other important factors, which this work builds upon, are  the existence  of three complete, well-defined samples of CSOs, and the fact that in some cases it is possible to assess the possible effects of selection biases on the topics under study. This is what we have attempted  in these three papers.

This paper is structured as follows:
\Cref{sec:criteria} introduces and discusses the \ac{cso} classification criteria.
In \cref{sec:selection} we describe our candidate selection and classification process and present the results of the classification. In \cref{sec:biases} we discuss the biases in our sample selection process.
\Cref{sec:samples} presents the fraction of \acp{cso} in complete statistical samples.
\Cref{sec:csos-angular-sizes-and-spectra} focuses on the  angular sizes and spectra of \acp{cso} and \cref{sec:csos-linear-sizes} on the physical sizes. Our conclusions are presented  in \cref{sec:conclusions}.
 We discuss other important aspects of CSOs in Appendix~\ref{app:mwl}, and the Table of 79 bona fide CSOs with references to the origins of the quoted numbers is given in  Appendix~\ref{app:csocat}.

% conventions:
Throughout this paper we adopt the convention $S_\nu \propto
\nu^{\alpha}$ for spectral index $\alpha$, and use the cosmological
parameters $\Omega_m = 0.27$, $\Omega_\Lambda = 0.73$ and $H_0 = 71 \;
\mathrm{km\; s^{-1} \; Mpc^{-1}}$ \citep{2009ApJS..180..330K}.

%------------------------------------------------------------------------------
\section{Revised CSO Classification Criteria}
\label{sec:criteria}

The \ac{cso} class was originally defined by W94 as: ``(1) two or more components, separated by 10--1000\,pc, either straddling a central core or with other compelling morphological evidence for symmetry such as outer hot spots, and (2) no emission on scales greater than 1000\,pc or only very faint emission.''  In addition,  P93 stated as the third property of CSOs, following the morphology and size, that ``the radio emission is not strongly beamed'', and  W94 emphasized that ``relativistic beaming plays at most a minor role in CSOs'', and in their summary they stated that ``relativistic beaming does not play a major role in determining the properties of CSOs''.  It should be clear, therefore, that blazars should never be classified as CSOs because blazars are heavily affected by relativistic beaming towards the observer.

Because their jets are closely aligned with the line of sight,   BL~Lac objects and \acp{fsrq}  can contaminate the  \ac{cso} class in one of two ways: (i) in some cases a relativistically boosted nucleus and a relativistically boosted jet component present the appearance of a \ac{cd}, (ii) in the case of some slightly  curved jets, different parts of the approaching jet are projected on opposite sides of the nucleus.

 R96   listed a dozen properties of \acp{cso}, based on five \acp{cso}  they had observed in the Pearson-Readhead complete sample \citep{1981ApJ...248...61P}. Unfortunately these properties have often been ignored in the subsequent literature, and this has contributed substantially to the mis-classification of \acp{cso}. 
 Of particular relevance here is property \#8 listed by R96: ``\acp{cso} exhibit only weak radio variability, up to about 10\% of the total flux density over periods of a few years''. 

Thus,  mis-classified \acp{cso} can frequently be  identified through observations of variability and, in some cases, component speed. This demonstrates the need for long-term radio monitoring of AGN, such as the Owens Valley Radio Observatory 40 m Telescope program \citep{2011ApJS..194...29R}.   With the addition of these requirements to those of \citet{1994ApJ...432L..87W}, we define four criteria, which establish the basis for compiling a comprehensive and definitive catalog of CSOs free from sources which are significantly affected by relativistic beaming: 

\begin{enumerate}
    \item No projected radio structure larger than 1\,kpc, with exceptions only in the case of repeated activity, as discussed below. 
    \item Evidence of emission on both sides of the center of activity. The latter may or may not be  identified, but its location should be well established through clear evidence of jets  and/or hotspots, and/or lobes.
    \item  The source should not be known to have a fractional variability  of greater than 20\%/yr.
    \item No  known superluminal motion in any jet component in excess of $v_\mathrm{app} = 2.5 \mathrm{c}$
\end{enumerate}
  Because it takes many years to characterize the variability of an AGN, and likewise to measure the speed of any moving components, not all clear-cut CSO candidates can currently be tested against all four criteria.  Nevertheless given the intense scrutiny to which candidates have been subjected (see Section 3) we assume that any of the 79 “bona fide” CSOs on which the evidence on variability and speed is currently lacking, will be proven to satisfy criteria 3 and 4.

\begin{figure*}[!t]
    \centering
    \includegraphics[width=1.0\linewidth]{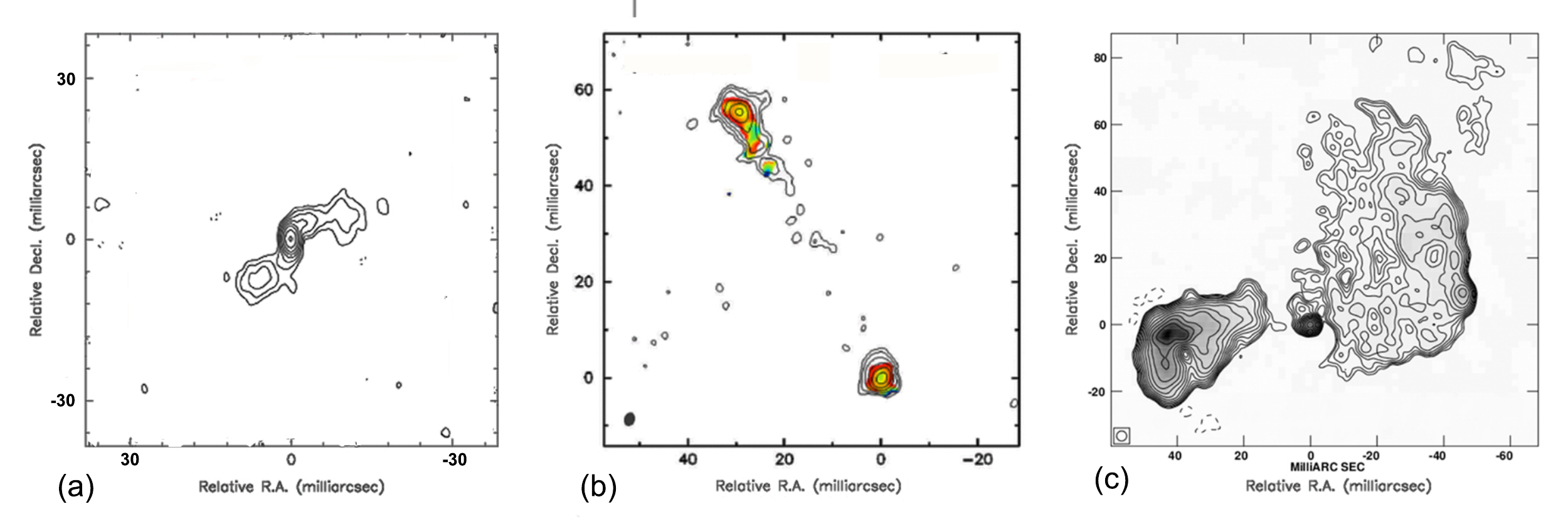}
    \caption{
        Three examples of bona fide \acp{cso} that pass our  selection criteria. (a) \ac{vlba} map of  J1220+2916 (NGC~4278) at 4.845\,GHz \citep{2007ApJ...658..203H} -- a \ac{cso}~1 object showing edge-dimmed morphology. The  peak flux density is $0.1208\,\mathrm{Jy}\; \mathrm{beam}^{-1}$.
        (b) \ac{vlba} map of J1159+5820 at 5\,GHz \citep{2016MNRAS.459..820T} -- a  \ac{cso}~2 object, in which the emission is edge-brightened and dominated by the hot spots and narrow jets. The contour levels begin at $1\;{\rm mJy \; beam^{-1}}$  and increase by powers of~2. 
        (c) \ac{vlba} map of J0119+3210 (B2\,0116+31) at 5\,GHz \citep{2003AA...399..889G}, a  \ac{cso}~2 object showing highly resolved edge-brightened morphology with faint hot spots close to the extremity of the source. The contours are $-0.4, 0.4, 0.57, 0.8, \dots, 36.2\,\mathrm{mJy}\,\mathrm{beam}^{-1}$. Negative contours are dashed. The peak flux density is $40.2\,\mathrm{mJy}\;\mathrm{beam}^{-1}$.
    }
    \label{fig:morphology}
\end{figure*}

Since 2008 we have been monitoring the 15 GHz flux densities of $\sim$1800  jetted-AGN $\sim$twice a week \citep{2011ApJS..194...29R} on the 40 m Telescope of the Owens Valley Radio Observatory (OVRO).  We therefore use 15 GHz as the fiducial frequency at which we estimate the variability of CSOs.
We have 15 GHz light curves of  10 of the 79 bona fide CSOs in our  sample, on which we have carried out a detailed study, and measured the maximum fractional variability per year.  We examined each light curve by eye, and selected the feature with the steepest slope.  We measured the time interval, $\Delta$t, between the time at the highest flux density, $t_{\rm hi}$, and the time at that the lowest flux density, $t_{\rm lo}$, and the change in flux density, $\Delta S= S_{\rm hi}- S_{\rm lo}$ between these, where $S_{\rm hi}$ and $S_{\rm lo}$ are the flux densities at $t_{\rm hi}$ and $t_{\rm lo}$, respectively.   We define the maximum fractional variability per year, allowing for cosmic time dilation, by $v_{\rm frac, max} = (\Delta S/S_{\rm lo})(1+z)/\Delta t$, where $z$ is the redshift of the object. The duration of the fastest varying  features in blazars in the 40 m monitoring sample ranges from a week, or less, up to 1.5 years, while in  the 10 CSOs for which we have light curves, the duration ranges from $\sim$1 year to $>$15 years.

 The OVRO light curve of the CSO OQ 208 (Fig. \ref{fig:lightcurve}), which is one of the nearest and most strongly and rapidly  varying CSOs, shows a steady decline in flux density, giving a maximum fractional variation rate of  $v_{\rm frac, max}=-11.1 \%$/yr. The strongest maximum fractional variation rate in  the ten CSOs monitored at the OVRO is seen in J1735+5049, with $v_{\rm frac, max}=14.8\%$/yr. For our purposes, we adopt a maximum fractional variability rate of $v_{\rm frac,\, CSO \, max}=20\%$/yr as an upper limit for CSO variability.  We have now undertaken a study in which we will quantify more comprehensively  the variability of all of the currently known bona fide CSOs at 15 GHz  over the next decade.

 For comparison with the above numbers for CSOs, it is interesting to note that the maximum fractional variation rate of PKS 1413+135 (Fig. \ref{fig:lightcurve}) is 808.9\%/yr, and the the fractional variation of the least variable object that we rejected in our  filtering step was 24\% per year, i.e. more than a factor 1.6 greater than the largest fractional variation rate we have seen in a CSO thus far.

We adopt an upper apparent velocity cutoff of $v_\mathrm{app} < 2.5 \,{c}$ for bona fide \acp{cso}. Higher apparent speeds would imply a jet viewing angle smaller than $45^\circ$ and a Lorentz factor greater than~3, and non-negligible relativistic beaming effects of order unity or higher. For this reason we reject any AGN as a CSO candidate if the apparent speed of any jet feature in the source exceeds the $2.5\,{c}$ speed limit.
 At this stage, these values are chosen based on experience. However, clearly what is needed here is a study of a large number (we suggest about 50) of CSOs in a complete flux density limited sample.  These objects must be identified and observed at multiple wavelengths for at least a decade in order to determine the relationship between the jets, the lobes, and the envelopes.  Only at that point will it become possible to be more precise about the beaming factors and the speeds that  obtain in CSOs.

We discuss these criteria further in \crefrange{sec:criteria-size}{sec:criteria-speed}. 
Three examples of \acp{cso}~1s and \acp{cso}~2s that pass  our vetting criteria are shown in \cref{fig:morphology}.

%------------------------------------------------------------------------------
\subsection{The Largest Projected Linear Size of CSOs}
\label{sec:criteria-size}

\citet{1995AA...302..317F} and R96 divided jetted-\acp{agn} into three classes: \acp{cso} with projected sizes less than 1\,kpc, \acp{mso} with sizes between 1\,kpc and 20\,kpc, and \acp{lso} with sizes greater than 20\,kpc.  These limits correspond roughly to the boundary between the  region dominated by the supermassive black hole and that dominated by the galaxy, and the boundary between the region dominated by the galaxy and that  dominated by the extragalactic environment.

Some \acp{cso} show clear evidence of a previous epoch of activity \citep{1990AA...232...19B}, in which a new epoch of activity   has begun to form  a \ac{cso}  while vestiges of a previous epoch of activity are still visible. In these \acp{agn} the radio surface brightness either drops off significantly or is completely absent in the region between the ($<1$\,kpc-scale) \ac{cso} and the outer structures.  In such cases the symmetric structure  on the  sub-kpc scale is used to define the size required for inclusion in the \ac{cso} class.

%------------------------------------------------------------------------------
\subsection{The Radio Morphology of CSOs}
\label{sec:criteria-morphology}

The symmetry \ac{cso} criterion includes objects which show {\it any} emission on both sides of the center of activity, or nucleus. The center of activity is identified either by a compact flat spectrum component, or clear symmetry of the lobes. It is not necessary that flux densities of the components straddling the nucleus be comparable. For example, an \ac{agn} in which the lobes straddling the nucleus have a  flux density ratio of 10:1, which passes the other CSO criteria, should be classified as a bona fide \ac{cso} according to our definition of a CSO.

Sometimes the center of activity cannot be clearly identified, so we do not require identification of the nucleus for inclusion in the \ac{cso} class if other morphological evidence makes it clear that the nucleus lies between the two outer structures.  
\vskip 6pt
\noindent
 Examples of other morphological evidence might include, but are not limited to, cases where:
\vskip 6pt
\noindent
(i) there are two clear, oppositely directed, steep spectrum jets with flat spectrum hot spots at one, or both of, the outer leading edges of the jets,
\vskip 6pt
\noindent
(ii) there are two clear, steep spectrum lobes,  with  flat spectrum hot spots at one or both of the outer edges of the lobes, and
\vskip 6pt
\noindent
(iii) there are two clear, steep spectrum lobes without any discernable hot spots.

%------------------------------------------------------------------------------
\subsection{Radio variability in CSOs}
\label{sec:criteria-variability}

 Since we do not have OVRO light curves for all CSO candidates, the $<20\%$/yr  variability   limit discussed above cannot be applied to all CSO candidates.    We therefore applied different approaches based on the available data,  as follows:
We rejected \acp{agn} from the \ac{cso} class whose radio light curves exhibit rapid, blazar-like flares, or whose multi-epoch radio spectra indicate variability over a wide range of frequencies in less than the light travel time across the source.
Additionally, we considered variability index values from the literature \citep[e.g.,][]{2006MNRAS.370.1556B,2010MNRAS.408.1075O}.

 As should be clear from the opening  paragraph of \S \ref{sec:criteria}, the absence of relativistic beaming, and hence of strong variability, should {\it always\/}  have been applied when classifying objects as CSOs. Unfortunately this has not been the case. Two examples of \ac{agn} that have been classified as \acp{cso} in the literature that we reject as \acp{cso} on the basis of their variability are discussed in detail in the following two subsections.  These are the unusual \ac{agn} PKS~1413+135 \citep{2021ApJ...907...61R}, and PKS~1543+005, another strongly variable radio source \citep{1981AJ.....86.1604D}.   PKS~1543+005 provides a prime example of projection of an approaching jet onto both sides of the nucleus, thereby masquerading as a \ac{cso} morphology because the emission does not come from two oppositely-directed jets.

\begin{figure*}[!t]
    \centering
    \includegraphics[width=1.0\linewidth]{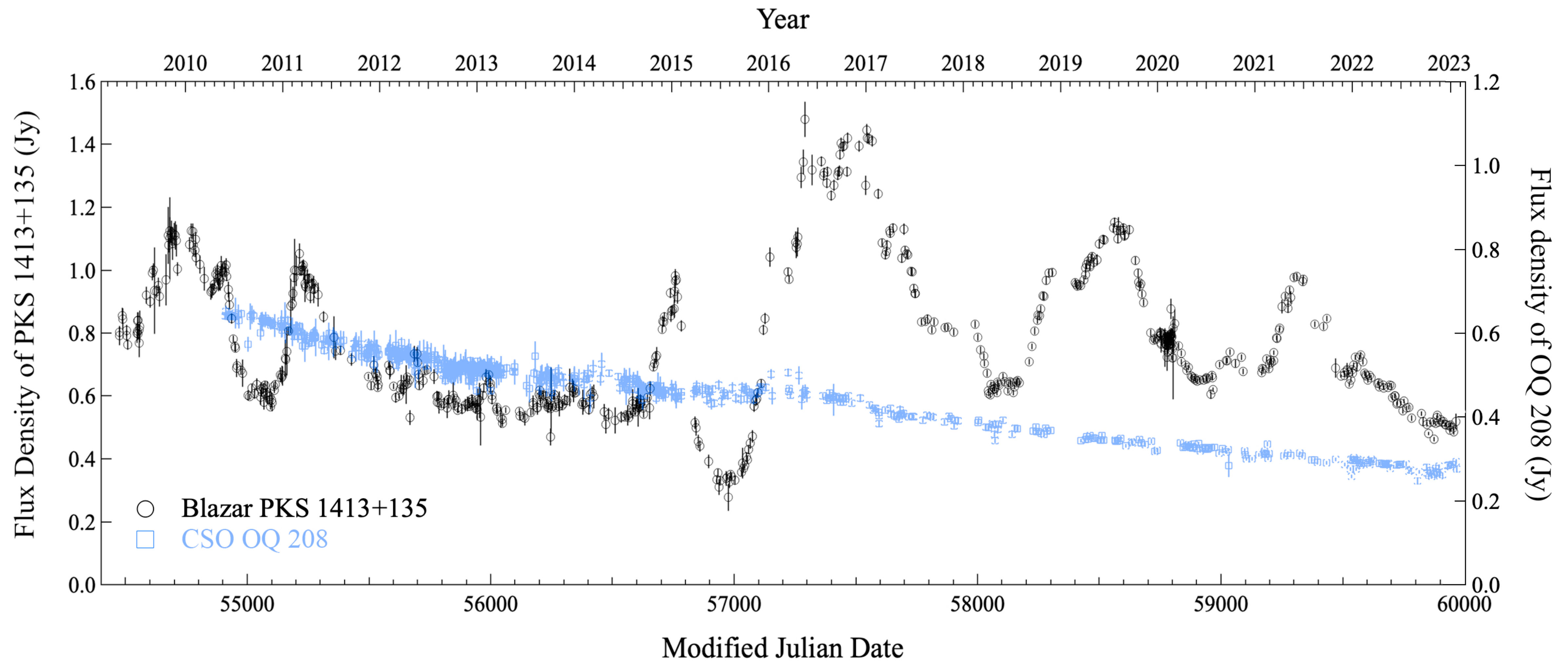}
    \caption{ OVRO 15 GHz light curves of a typical blazar and a typical CSO.  PKS 1413+135 (black points) is a prime example of a blazar that should  not have been classified as a CSO, but which is nevertheless  much discussed  in the literature as a CSO. The mis-classification is obvious in this comparison of the  PKS 1413+135 light curve with the light curve of OQ 208 (blue points), which is one of the most rapidly varying CSOs. The maximum fractional variation in OQ 208 is $v_{\rm frac, max}=-11.1 \%$/yr, whereas the maximum fractional variation in PKS 1413+135 is $v_{\rm frac, max}=808.9 \%$/yr (see text).
    }
    \label{fig:lightcurve}
\end{figure*}

%------------------------------------------------------------------------------
\subsubsection{PKS~1413+135}
\label{sec:pks1413}

PKS~1413+135 has been discussed as a \ac{cso}, or CSO candidate, in many studies \citep[e.g.,][]{1994ApJ...424L..69P,1996AJ....111.1839P,2002AJ....124.2401P,2005ApJ...622..136G,2010ApJ...713.1393W,2021MNRAS.507.4564P}.  In Fig. \ref{fig:lightcurve}  we show its  15 GHz light curve. This object clearly violates the fundamental requirement for classification as a CSO: that the observed radiation should not be strongly beamed towards the observer (R93, R94, W94, R96).  The variability of PKS 1413+135  as revealed by the light curve from the  OVRO monitoring program, makes it obvious that this object is a blazar. PKS 1413+135 is one of the most strongly varying jetted-AGN in the OVRO monitoring program.   Early evidence of very strong variability at infrared wavelengths was reported by \citet{1981Natur.293..714B}, who reported that PKS 1413+135 is ``among the most highly variable extragalactic sources known'' and that at 2.2 $\mu$m it exhibited changes of $>10\%$ on timescales of 1 day, and on three occasions the intensity changed by over a factor of two in less than 1 month.
 
\begin{figure*}[!t]
    \centering
    \includegraphics[width=1.0\linewidth]{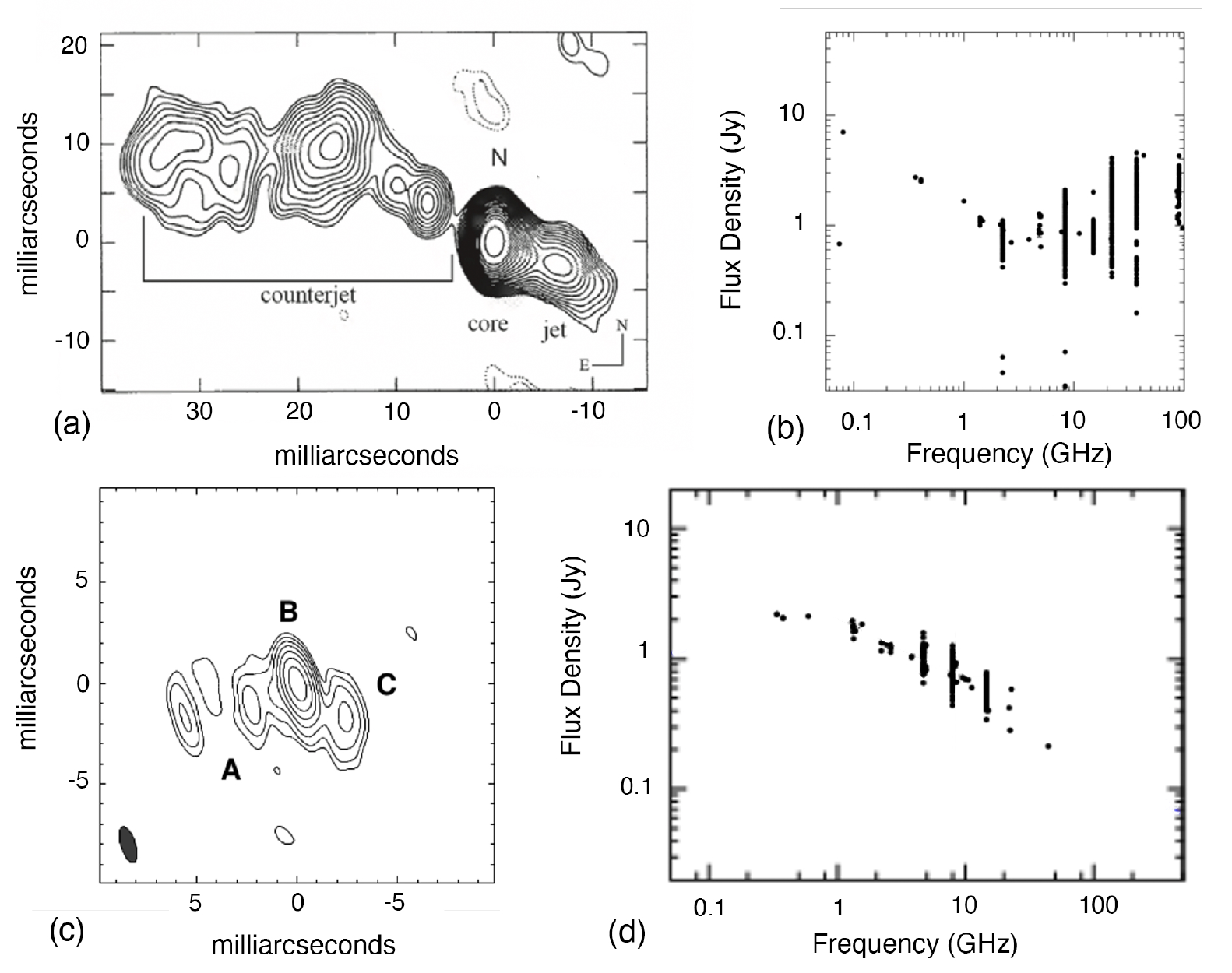}
    \caption{Two examples of jetted-\acp{agn} that have been misclassified as \acp{cso} in the literature. These have now been rejected since their variability is  $> 20\%$/yr (see text). (a) \& (b) PKS~1413+135, (c) \& (d) PKS~1543+005. (a) VLBA map of PKS~1413+135 at 5\,GHz \citep{1996AJ....111.1839P} showing clear emission in a jet and counter-jet that straddle the flat-spectrum radio nucleus at ``N''. The contour intervals are at $6 \times 10^{-4}\,\mathrm{Jy\,beam}^{-1}$. (b) the variability of PKS~1413+135 at different frequencies is shown here in this plot of flux densities taken from the MOJAVE website \citep{2019ApJ...874...43L}.  The points at each frequency represent the range of total flux densities for this source observed at each frequency. (c) The morphology of PKS~1543+005 in this 15\,GHz image from \citet{2000ApJ...534...90P} apparently shows structure from a jet and a counter-jet straddling the flat-spectrum nuclear component at ``B''. However, the distance between ``C'' and ``B'' is decreasing over time.  The peak flux density is $289.4 \; {\rm mJy}\; {\rm beam}^{-1}$.  As seen in (d) (from \citealt{2008AA...482..483T}) this is a highly variable \ac{agn}, making this an example of a slightly bending jet moving towards us almost along the line of sight that is being projected onto both sides of the nucleus (see text). }
     \label{fig:noncsos}
\end{figure*}

All of this  demonstrates clearly that this object is a blazar, and therefore not  a CSO.  In \cref{fig:noncsos} we show its morphology and its spectrum including variability. Morphologically, PKS~1413+135 meets the criteria for a \ac{cso} since there is no doubt that it shows both a jet and a counter-jet on these small scales, and it is less than a kpc in total extent. However it is a blazar and a BL~Lac object, and its high variability, at both radio and infrared wavelengths, as discussed in  detail by \citet{2021ApJ...907...61R},  leaves no doubt that its axis is closely aligned with the line of sight and that its nuclear emission, as well as that in the approaching jet on the south-west side of the nucleus, is relativistically boosted towards us.   It is  clear from the case of PKS 1413+135 that the fact that the CSO classification was defined specifically to exclude objects in which the observed emission is strongly relativistically boosted towards the observer (R93, W94, R96) has not been widely recognized.

%------------------------------------------------------------------------------
\subsubsection{PKS~1543+005}
\label{sec:pks1543}

PKS~1543+005 was classified as a \ac{cso} by \citet{2000ApJ...534...90P}. In  \cref{fig:noncsos}  we show its morphology and its spectrum
including variability. Like PKS~1413+135, morphologically PKS~1543+005 resembles a \ac{cso} since there is no doubt that it shows emission on both sides of the nucleus (component ``B'') on these small scales, and it is less than a kpc in total extent. Like PKS~1413+135, it is a highly variable blazar, as shown by \cite{1981AJ.....86.1604D}. PKS~1543+005 also has a spectral peak luminosity over three times greater than any of the 79~\acp{agn} we identify as bona fide \acp{cso} in this paper.  Component ``C'' is apparently moving towards component ``B'' with apparent speed $(1.10 \pm 0.17)c$ \citep{2005ApJ...622..136G}.  These authors point out that this could indicate that component ``C'' is stationary, while component ``B'' is apparently moving towards component ``C'',  due to the ejection of a new nuclear component.  However, the variability shown in \cref{fig:noncsos}~(d) is a factor of 4 at 10\,GHz.  Furthermore, in a study of 90~\acp{agn} at 318\,MHz, \citet{1981AJ.....86.1604D} found PKS~1543+005 to be by far the most variable of the \acp{agn} in their sample.  Their conclusion about their most variable objects was  that ``these properties are consistent with models invoking relativistic beaming''. On account of its high variability and high luminosity there can be no doubt that the radio emission from PKS~1543+005 is highly boosted by relativistic beaming, and thus this object should  not be classified as \ac{cso}.

Thus PKS~1543+005 likely falls into one of the two cases of misidentified \acp{cso} that we are trying to avoid in the \ac{cso} class: those closely aligned with the line of sight whose apparent symmetry is due to approaching structure being projected
onto both sides of the nucleus.

\begin{figure*}[!t]
    \centering
    \includegraphics[width=1.0\linewidth]{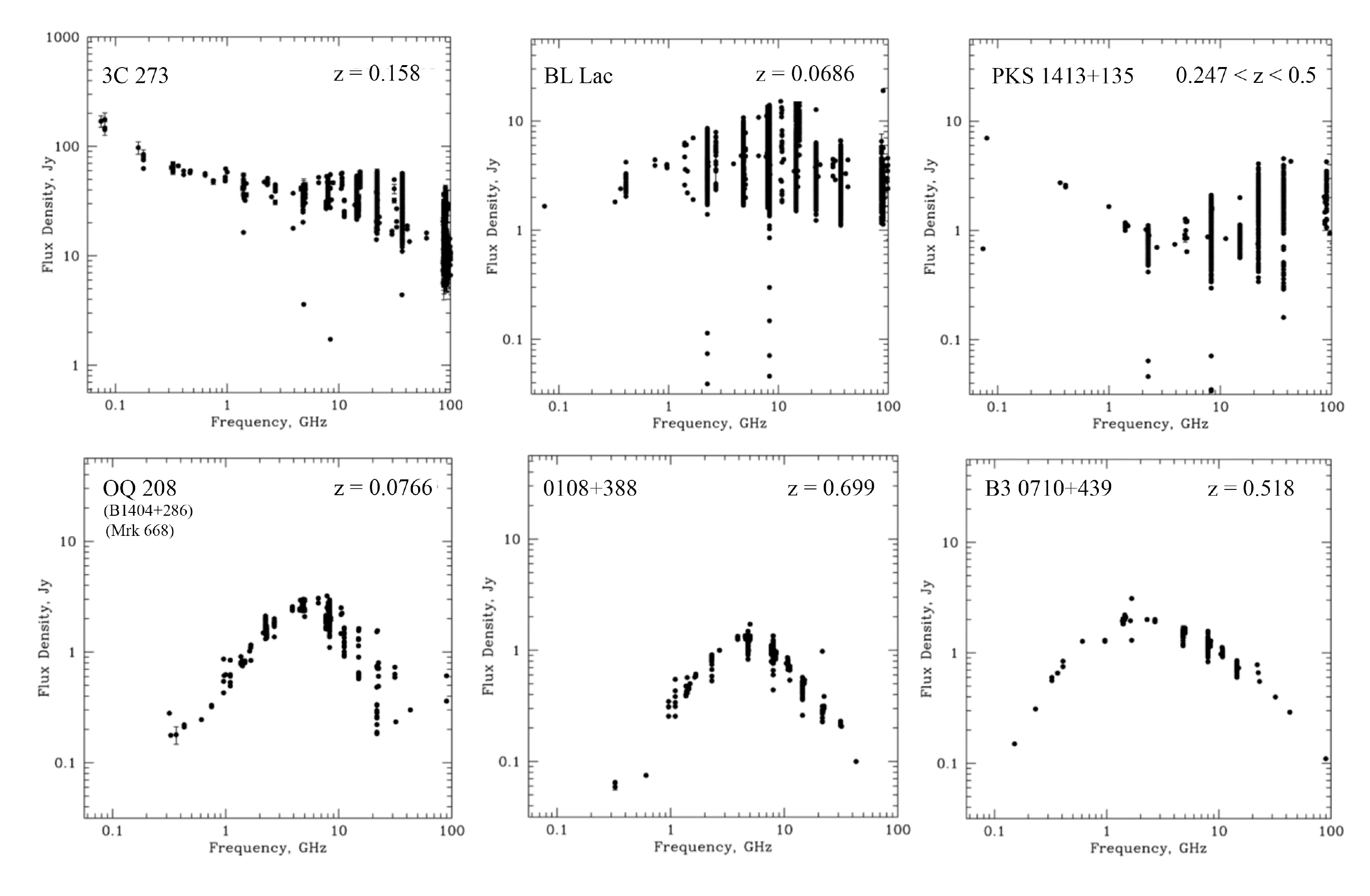}
    \caption{ The spectra and variability of Blazars {\it vs.\/} CSOs. The top three panels show the spectra and the variability of three blazars: 3C 273, a flat spectrum radio quasar (FSRQ); Bl Lac, the archetypal BL Lac object; and PKS 1413+135, an object that has frequenctly been misclassified as a CSO, whose 15 GHz light curve is shown in \cref{fig:lightcurve}. The bottom three panels show the spectra and variability of three typical CSOs.  These peaked spectrum sources are typical of CSOs, although $\sim 10\%$ of the CSOs in our bona fide sample have monotonically rising spectra towards low frequencies. CSOs show only slow variations in both flux density and structure (see text). 
    }
    \label{fig:specvar}
\end{figure*}

\vskip 6pt
\noindent

{\it Summary of CSO variability filter:} Any misconception regarding the rate of variability in CSO spectra and structure,  can be resolved using the variability data of the CSOs in the OVRO 15 GHz monitoring program. Typical examples of the spectra and variability of CSOs are shown in the lower three panels of Fig. \ref{fig:specvar}, where they are compared to those of three blazars, shown in the upper three panels.   Although the fractional variations in these CSOs are large,  as shown in Fig. \ref{fig:specvar}, they occur over a long time period. A prime example is that of OQ 208, shown in \cref{fig:lightcurve}.

%------------------------------------------------------------------------------
\subsection{Apparent Velocities of Components in CSO Jets}
\label{sec:criteria-speed}

In general, the measured apparent speeds of components in the jets of \acp{cso} are found to be $v_\mathrm{app}\lesssim c$ \citep{ 2016AJ....152...12L, 2016MNRAS.459..820T}.
The mean separation speeds of the hot spots are found  on average to  be $\sim 0.36\,{c}$ \citep{2000ApJ...541..112T}, \citep{2002evn..conf..139P, 2012ApJ...760...77A}, and Paper~2; but some jet components in \acp{cso} have speeds close to, or exceeding, $c$.
For example the maximum apparent speeds of components in the jets of three typical \acp{cso}  are J0111+3906 (0108+388): $v_\mathrm{app,max} = (0.83 \pm 0.15)\,{c}$; J0713+4349 (B3 0710+439): $v_\mathrm{app,max} = (1.03 \pm 0.32)\,{c}$  \citep{2019ApJ...874...43L}; and  J1945+7055: $v_\mathrm{app,max} = (1.088 \pm 0.011)\,{c}$ \citep{2009ApJ...698.1282T}.

\section{The Search for CSO Candidates and the Selection of 79~Bona Fide CSO\lowercase{s}}
\label{sec:selection}

The principal objective of this paper is to define and apply the above revised \ac{cso} selection criteria in order to filter out \acp{agn} incorrectly identified as \acp{cso} in the literature, and hence to lay a firm foundation for studies of the phenomenology of \acp{cso}.   

We therefore undertook an extensive literature search for sources that had been identified as \acp{cso} or potential \ac{cso} candidates. We augmented our initial list of candidates with other \acp{agn} that we thought might well be \acp{cso}, as well as all \acp{agn} from three complete flux density-limited samples. We then examined all of these sources in the light of our four \ac{cso} selection criteria.

It is not our intention in this paper to present a large complete catalogue of \acp{cso}.  We hope that the  sample  of bona fide \acp{cso} we present will be added to and developed by the community into a  catalog of bona fide \acp{cso}.

In \crefrange{sec:search}{sec:vetting} we describe our literature search and the vetting process.

%------------------------------------------------------------------------------
\subsection{CSO candidate search}
\label{sec:search}

% literature search:
We embarked on a literature search for \ac{cso} candidates
using the \verb|astroquery|\footnote{\url{https://astroquery.readthedocs.io/en/latest/}} \citep{2019AJ....157...98G} interface to the \ac{ads}\footnote{\url{https://ui.adsabs.harvard.edu/}}. We searched for refereed publications that contained the term ``compact symmetric object'' in the title or abstract with the query \verb|abs:(compact symmetric object)| \verb|property:refereed database:astronomy|.
From the query results we filtered 125~publications that explicitly mentioned ``compact symmetric object'' in the abstract.
Similarly, we found 490~publications using the search terms ``compact steep spectrum'' and ``gigahertz peaked spectrum'', out of which we identified 17 containing \ac{vlbi} images of \ac{cso}-like \acp{agn}. Several dozen other publications were subsequently added during the literature review, leading to a final list of approximately 200~publications.

% other compilations:
In the $\sim 30$~years since the original definition paper for the \ac{cso} class by W94, there have been numerous papers that have presented lists of \acp{cso}, peaked radio sources, and young radio jets.
\cite{2006MNRAS.368.1411A} adopted a  systematic approach and literature search similar to ours. They compiled a sample of 37~\acp{cso}, and 4~candidate \acp{cso} with unknown redshift.
\cite{2006MNRAS.368.1411A} also constructed a list of 157~candidate sub-kpc-sized flat-spectrum \acp{cso}, of which they ruled out 61 as~\acp{cso}.
\cite{2007AA...463...97L} expanded the original list of \ac{ps} sources by \cite{1991ApJ...380...66O} to create a new master list of 74~\ac{ps} radio sources.
\cite{2008AA...482..483T} collected data on a much larger sample of 206~\acp{ps} and \acp{hfp} from the literature. Their analysis confirmed a strong contamination of the \ac{hfp} class by blazars which have temporarily peaked spectra during periods of radio flaring. 
\cite{2012ApJ...760...77A} presented a compilation of known \acp{cso} and \ac{cso} candidates, which included six sources that \cite{2012ApJS..198....5A} classified as new \acp{cso}. 
\cite{2014MNRAS.438..463O} constructed a list of 51~young radio sources that had unambiguous core region detections, of which 19 are smaller than 1\,kpc. 
\cite{2016MNRAS.459..820T} performed multi-frequency \ac{vlba} follow-up observations of 109~\ac{cso} candidates in the \ac{vips} and identified nine previously confirmed and 15~new \acp{cso}. There is considerable overlap in the \ac{cso} source lists from these compilations.
All of these source compilations have been considered in our source sample.

% complete samples:
We augmented the \ac{cso} candidate source list with all sources from the following complete radio \acp{agn} samples: the \ac{pr} sample \citep{1988ApJ...328..114P}, the \ac{cj1} sample \citep{1995ApJS...98....1P}, 
the 171~sources from the \ac{pw} sample \citep{1981MNRAS.194..331P,1985MNRAS.216..173W} with flux density S$_{\rm 2.7 \,\,GHz} \geq 1.5\,\mathrm{Jy}$.   In addition we used the incomplete
 \ac{cjf} sample \citep{1996ApJS..107...37T}, and the \ac{vips} sample \citep{2007ApJ...658..203H}. Note that the original PW sample \citep{1981MNRAS.194..331P} consisted of 168 objects, to which three more (DA~240, 0945+73 = 4C~73.08, and NGC~6251)
 were added by \citet{1985MNRAS.216..173W}.
% final list and database:
This process yielded a list of 3,175~\acp{agn} that were potential \ac{cso} candidates. We assigned a unique ID number to each of these candidates.
We compiled a database of these sources, which
includes (where available): angular sizes,  radio spectra, jet component speeds, radio light curves, variability indices, and polarization. We used only spectroscopic redshifts, both from the literature, and from  the \ac{ned}\footnote{\url{http://ned.ipac.caltech.edu/}}.

VLBI images obtained from the literature were complemented by images and multi-epoch radio spectra from the online \ac{rfc}\footnote{\url{http://astrogeo.org/rfc/}} and also VLBI images from the MOJAVE program archive\footnote{\url{http://www.cv.nrao.edu/MOJAVE/}} \citep[e.g.,][]{2019ApJ...874...43L}.

For 1,657 of the sources in our list, 15\,GHz light curves were available from the \ac{ovro} 40\,m monitoring program\footnote{\url{https://sites.astro.caltech.edu/ovroblazars/}} \citep{2011ApJS..194...29R}. We used these to check whether the light curves showed flares typical of blazars.

%------------------------------------------------------------------------------
\subsection{CSO vetting process}
\label{sec:vetting}

We classified all 3,175~sources as follows,   see Table \ref{tab:classification}, which gives the relevant numbers:

\begin{description}
    \item[Bona fide \acp{cso}] Sources that were not rejected on the basis of any of our \ac{cso} criteria. 
    \item[A-class candidates] Sources that were not rejected by the \ac{cso} criteria, but for which the available \ac{vlbi} images do not definitively confirm or rule out their \ac{cso} nature.  We are obtaining new \ac{vla} and/or \ac{vlba} observations  of these \acp{agn}, which we consider to be promising \ac{cso} candidates (in preparation).
    \item[B-class candidates] These sources were not rejected on the basis of size, speed, or variability measurements, but lacked sufficient \ac{vlbi} data to confidently classify their morphology as  \acp{cso}. Roughly 40\% of these sources have no available \ac{vlbi} images,  and another $\sim 30\%$ are unresolved in existing \ac{vlbi} images.  We cannot rule out these ``B-class'' sources as potential \acp{cso}, but considered them to be less promising \ac{cso} candidates for follow-up \ac{vlba} studies, due to  low flux density, unresolved structure, and/or location in the southern sky. The large number of possible \acp{cso} in this class (\cref{tab:classification}) strongly suggests that any compilation of \acp{cso} at this stage that is not very carefully constrained by the selection criteria is likely to be seriously incomplete. 
    \item[Rejected candidates] These sources were rejected on the basis of at least one of the \ac{cso} classification criteria described in \cref{sec:criteria}.
\end{description}

% automated classifications:
We applied an initial filter by rejecting all \acp{agn} in our source list with  projected largest  size exceeding 1\,kpc, except for five \acp{agn} that showed evidence of more than one epoch of activity, which are discussed below. All five of these \acp{agn} showed an emission gap, or a strong decrease in surface brightness, between the inner ($< 1\,\mathrm{kpc}$) and outer (kpc-scale) radio structures. 
Most of the \ac{css} sources in our list were filtered out in this step. For \acp{agn} without redshifts, we rejected all those with angular size greater than 1\,arcsecond.  We note that this may have potentially excluded some low redshift \acp{cso} ($z < 0.06$).  

Of our candidates, 228~were monitored in the MOJAVE program \citep[e.g.,][]{2019ApJ...874...43L}, which provides measurements of the radio component speeds. We rejected 144~sources from our list that had measured speeds exceeding $2.5\,{c}$.

% manual vetting process:
The remaining candidates were examined independently by at least two of our co-authors, at least one of whom had significant experience in \ac{vlbi} and the study of relativistic jets. Most of the rejected sources were obvious asymmetric, one-sided ``core-jet'' sources that could  therefore be rejected immediately on the basis of their morphology.
A few could be classified immediately as ``bona fide'' \acp{cso}. Any candidate deemed a possible \ac{cso}, but not definite, by one of the examiners, was discussed by the whole group.
Finally we assigned all the A- and B-class sources to one author each to gather more data from the literature, if available, and to confirm the prior classification or suggest another round of group discussion, where a final classification was determined by consensus.
Inevitably there were marginal classification choices, and we accept that there will be sources which turn out to be \acp{cso} amongst the B-candidates that we decided not to follow up.

% VIPS:
The sample of \ac{vips} sources was added at a later stage. A fraction of the sample had already been classified by us. For the newly added \ac{vips} sources we (i) automatically rejected those that had been classified as core-jets by \citet{2016MNRAS.459..820T}, (ii) automatically classified those as B-class candidates that were identified as point sources by \citet{2016MNRAS.459..820T}, and (iii) examined those sources identified by \citet{2016MNRAS.459..820T} as \acp{cso} or complex, in the same manner as described above.

% TABLE: classification
\begin{deluxetable*}{lrl}
    \tablecaption{
        Number of sources in each group.
        \label{tab:classification}
    }
    \tablehead{
        \colhead{Group} & \colhead{Count} & \colhead{Comments}
        }
    \startdata 
                CSO &     79& 15 newly confirmed bona fide CSOs (this paper). These new confirmed CSOs are indicated in Table \ref{tab:csos} \\
        A-candidate &    167&These  CSO candidates   are highly likely to be CSOs, which have been observed with new VLBA observations\\
        B-candidate &   1164&This large number of CSO candidates will require a large follow-up program\\
           Rejected$^\dagger$ &   1765& Grounds for rejection: morphology (1221), size (362), variability (194), speed (144)\\
           Total &   3175 &Claimed CSOs and CSO candidates\\
    \enddata
    \tablecomments{$^\dagger$  The numbers of CSO candidates rejected in this study for the reason(s) indicated in the third column. Some objects were rejected for more than one reason, but we did not determine all the reasons for rejecting a CSO candidate, usually stopping once a candidate had failed one criterion, since this is all that is needed to disqualify an object from the CSO class.}
\end{deluxetable*}

%------------------------------------------------------------------------------
\subsection{Classification}
\label{sec:classification}

% results:
\Cref{tab:classification} lists the total number of sources and the number of sources in each of the groups the sources were assigned to.
 The table also lists the number of sources that were rejected based on each criterion. 
A list of the 79~bona fide \acp{cso} is given in \cref{tab:csos} in Appendix~\ref{app:csocat} .
In this section we explain some specific classification decisions made in the context of the criteria defined above.

\subsubsection{CSOs Lacking an Identified Center of Activity}
\label{sec:nocenterofactivity}

As discussed in \cref{sec:criteria-morphology} it is not always possible to identify the center of activity from the available \ac{vlbi} images.
We classified as bona fide \acp{cso} 22~sources for which no center of activity has been detected, and 10 for which the core is not clearly identifiable.
These sources show compact double lobe-like features with steep spectra, sometimes with jet emission between them.

%------------------------------------------------------------------------------
\subsubsection{CSOs with Spectroscopic Redshifts}
\label{sec:noredshift}

Of the 79~bona fide \acp{cso} we have identified, only 54~bona  fide CSOs have published spectroscopic redshifts. These  are  shown in \cref{fig:redshiftdist}. The peak at low redshift is due to (edge-dimmed) \ac{cso}~1s, which are predominantly nearby, low luminosity \acp{agn}.  The  redshift distribution { can be affected by several selection biases that are discussed in \cref{sec:bias-size}.

% FIGURE: redshift distribution
\begin{figure}
    \centering
    \includegraphics[width=\columnwidth]{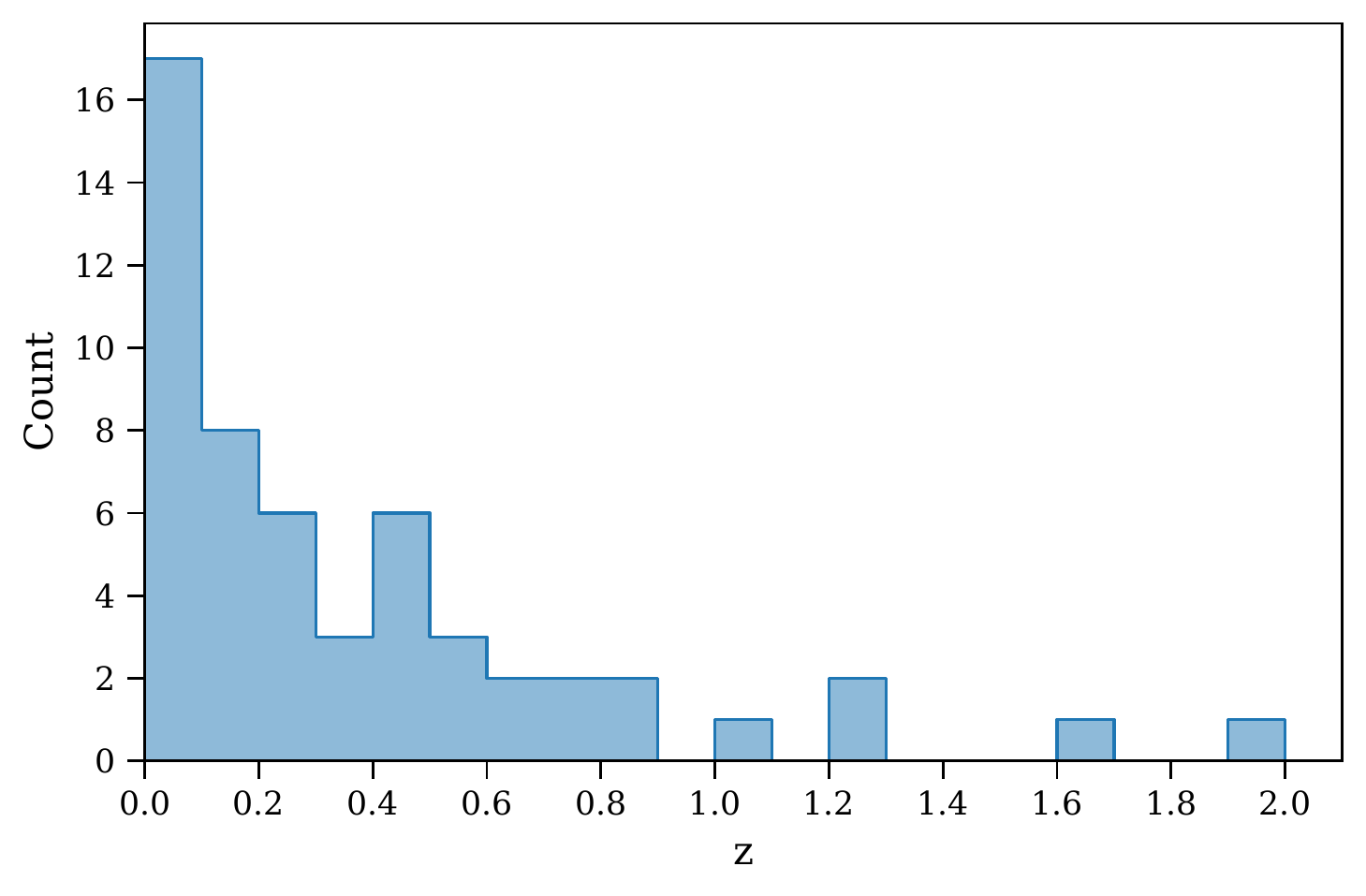}
    \caption{Redshift distribution of the 54~bona fide \acp{cso} with spectroscopic redshifts.   }
    \label{fig:redshiftdist}
\end{figure}

%------------------------------------------------------------------------------
\subsubsection{Long-term variability}
\label{sec:variability}

Of the 10 \ac{ovro} 40\,m CSO light curves, 8 show long-term trends in their 15\,GHz flux densities, with fractional variability from 10\% to 60\% on time scales from 5--15~years, but for all of them the maximum fractional variability rates are  $v_{\rm frac, max}<20\%$/yr.
As discussed in \cref{sec:criteria-variability}, slow variability is not in violation of criterion~3 we accept these sources into the class of \acp{cso}. The variability of \acp{cso} is an area of great interest.  The CSOs in our bona fide list for which we do not have OVRO light curves are not core-dominated and have very clear two-sided structure, so it is clear that they cannot be varying  strongly on timescales of a few years.

%------------------------------------------------------------------------------
\section{Sample Biases}\label{sec:biases}

Since the parent sample, from which we identified  79~sources as bona fide \acp{cso}, was accumulated from a large literature search, it is subject to a host of selection biases, the most severe of which we discuss below.

%------------------------------------------------------------------------------
\subsection{Random selection bias}
\label{sec:bias-random}

Some of our \ac{cso} candidates were selected from statistically ill-defined studies.  In many cases, high-sensitivity multi-frequency \ac{vlbi} observations revealed \ac{cso} morphology that might not be so readily apparent in a general broad \ac{vlbi} survey. For these \acp{cso} there is no way of estimating the selection biases.
It is important to keep this in mind, unless working with a particular well defined and carefully selected, sub-sample of the 79~bona fide \acp{cso} for which the selection effects can either be eliminated or do not affect the particular investigation being undertaken.

Since many of our \acp{cso} were discovered in surveys with well-defined selection criteria,  one can investigate potential selection biases, if any, in these cases.  We discuss these below.

%------------------------------------------------------------------------------
\subsection{Angular size and redshift bias}
\label{sec:bias-size}

There are two angular size selection effects in \ac{vlbi} surveys caused by the finite range of baseline lengths in the observing arrays.
The shortest baselines limit the lowest spatial frequencies, i.e., the largest extent of the  observable field of view. Long baseline interferometers, such as the \ac{vlba}, typically lack short baselines and are insensitive to components larger than $\sim 100\;\mathrm{mas}$ at 5\,GHz. 

On the other hand, the longest baselines determine the resolving power, and the finite resolution of cm-wave \ac{vlbi} means that \acp{cso} can only be identified down to lobe or hotspot separations of a few milliarcseconds. 

The most significant selection bias relevant to this study is that the  field-of-view limitation makes it easy  to miss  widely separated components at cm wavelengths unless one also has VLBI observations at a lower frequency.

It should be clear that the above angular size selection effects also impact the redshift distribution of our sample of bona fide \acp{cso}, because of the dependence of the apparent angular size on redshift.

%------------------------------------------------------------------------------
\subsection{Flux density limit bias}
\label{sec:bias-flux}

The  flux density limits of parent samples could bias against \acp{cso} having components with large angular  size and low surface brightness.

%------------------------------------------------------------------------------
\subsection{Spectral index limit bias} 
\label{sec:bias-spectral}

Spectral index limits can  bias the selection of \acp{cso} because  many CSOs have peaked spectra. Thus spectral index limits can exclude some CSOs.  In the CJF and VIPS samples, based on the results from our three complete samples, we estimate that $\approx 50\%$ of the CSOs above the flux density limit are missed due to the spectral index limit $\alpha \geq -0.5$.

%------------------------------------------------------------------------------
\section{Complete samples}
\label{sec:samples}

% TABLE: samples
\begin{deluxetable*}{llrrrrr}
    \tablecaption{
        CSOs and CSO candidates in three complete radio samples, and two samples that include only flat spectrum ($\alpha \geq -0.5$) sources down to a specified flux density limit, and are therefore incomplete. Rows~(2)--(8): the sample selection criteria. The corresponding references are listed in the table notes. Row~(9): the total number of sources in the samples. Rows~(10)--(13):the number of sources in each CSO category as defined in \cref{sec:vetting}. Row~(14): the fraction of \acp{cso} in each sample.
        \label{tab:samples}
        }
    \tablehead{
        \colhead{{\footnotesize (1)}} & & \colhead{PR} & \colhead{PR+CJ1} & \colhead{PW~1.5\,Jy} & \colhead{CJF} & \colhead{VIPS} \\
        &&(Complete)&(Complete)&(Complete)&(Incomplete)&(Incomplete)
    }
    \startdata
        {\footnotesize (2)} & Dec. & $\delta \geq 35^\circ$ $^*$ & $\delta \geq 35^\circ$ $^*$ & $\delta \geq 10^\circ$ $^*$ & $\delta \geq 35^\circ$ $^*$ & $65^\circ \geq \delta \geq 15^\circ$ $^{**}$ \\
        {\footnotesize (3)} & Galactic latitude & $|b|>10^\circ$ & $|b|>10^\circ$ & $|b|>10^\circ$ & $|b|>10^\circ$ & - \\
        {\footnotesize (4)} & Selection frequency & 5\,GHz & 5\,GHz & 2.7\,GHz & 4.85\,GHz & 8.5\,GHz \\
        {\footnotesize (5)} & Flux density & $S>1.3\,\mathrm{Jy}$ & $S>0.7\,\mathrm{Jy}$ & $S>1.5\,\mathrm{Jy}$ & $S>350\,\mathrm{mJy}$ & $S>85\,\mathrm{mJy}$ \\
        {} & {} & {} & {} & {} & {} & {} \\
        {\footnotesize (6)} & Spectral index & - & - & - & $\alpha^{4.85\,\mathrm{GHz}}_{1.4\,\mathrm{GHz}} \geq -0.5$ & $\alpha^{4.85\,\mathrm{GHz}}_\mathrm{low~freq} \geq -0.5$ \\
        {\footnotesize (7)} & Other constraints & - & - & - & - & Area of SDSS DR5 \\
        {\footnotesize (8)} & Reference & 1 & 2 & 3 & 4 & 5 \\
        \midrule
        % classification counts:
        {\footnotesize (9)} & Total & $^\dagger$64 & $^\dagger$199 & $^\dagger$170 & 293 & 1127 \\
        \midrule
        {\footnotesize (10)} & CSO & 6 & 12 & 13 & 11 & 33 \\
        {\footnotesize (11)} & Class A & 0 & 5 & 0 & 11 & 48 \\
        {\footnotesize (12)} & Class B & 1 & 4 & 4 & 29 & 268 \\
        {\footnotesize (13)} & Rejected & 57 & 178 & 153 & 242 & 778 \\
        \midrule
        % classification fractions:
        {\footnotesize (14)} & CSO fraction [\%]   & 9.4 &     6.0 &  7.6 &  3.8 &   2.9 \\
    \enddata
    \tablecomments{
        % references:
        $^*$ For source coordinates in B1950.0 equinox. $^{**}$ For source coordinates in J2000.0 equinox. $^\dagger$ We removed the starburst galaxy M82 from the sample. References: (1) \citet{1988ApJ...328..114P}, (2) \citet{1995ApJS...98....1P}, (3) \citet{1981MNRAS.194..331P,1982MNRAS.198..843P,1985MNRAS.216..173W}, (4) \citet{1996ApJS..107...37T}, (5) \citet{2007ApJ...658..203H}.
    }
\end{deluxetable*}

We searched for bona fide \acp{cso} in our three  complete radio samples, and  our two incomplete samples, which  excluded steep spectrum objects, i.e. those with $\alpha<-0.5$. The results are listed in \cref{tab:samples}.
The bona fide \ac{cso} detection fractions for these samples range
from 2.9\% to 9.4\%, but  both the CJF and VIPS samples are flat-spectrum $\alpha \geq -0.5$ samples and, as a result, are seriously incomplete, as explained in \cref{sec:bias-spectral}.

 In \cref{tab:samples} we see that in the \ac{pr} complete sample
 there are 6~bona fide \acp{cso}, and 0~Class~A candidates, and 1~Class~B candidate. For the \ac{cj1} sample these numbers are 12, 5, and 4, respectively, while for the \ac{pw} sample they are 13, 0, and 4, respectively.  If none of the Class~A and Class~B candidates in these complete samples turn out to be bona fide \acp{cso},  the fraction of \acp{cso} with $S_{\rm 5\,GHz}>700\,\mathrm{mJy}$  is $(6.8\pm 1.6)\%$. On the other hand, if all of the  Class~A candidates and none of the Class~B candidates turn out to be bona fide \acp{cso},  the fraction of \acp{cso} with $S_{\rm 5\,GHz}>700\,\mathrm{mJy}$  is $(8.5\pm 1.8)\%$.
 
 Thus  the fraction of \acp{cso} with  $S_{\rm 5\,GHz}>700\,\mathrm{mJy}$ lies between $(6.8 \pm 1.6)\%$ and  $(8.5 \pm 1.8)\%$.

%------------------------------------------------------------------------------
\section{The Angular Sizes and Spectra of CSO\lowercase{s}}
\label{sec:csos-angular-sizes-and-spectra}

In addition to the selection effects described above, our angular sizes were estimated using images at different frequencies, and for sources located at very different redshifts.
We therefore caution the reader not to over-interpret \crefrange{fig:redshiftdist}{fig:linsizedist}.

%------------------------------------------------------------------------------
\subsection{The Largest Angular Size of CSOs}
\label{sec:csos-angular-sizes}

In order to have as consistent a set as possible of measurements of the largest angular size,
we measured these  on the radio maps of the bona fide \acp{cso}, using the largest separation of the second lowest contour on the map because the lowest contour can often be noisy.  In cases where the component at the extremity of the map was unresolved, we used the position of the peak rather than the second contour. Whenever possible we used the lowest frequency \ac{vlbi} images available, as these are more sensitive to steep-spectrum low surface brightness emission. In some cases our size measurements are up to $\sim 15\%$ smaller than some published values that use different assumptions about the beam size and image noise level. We point out two large discrepancies: (i) B3\,0703+468 (J0706+4647), which \cite{2002AA...389..115D} measured as 75~mas, but we adopt a size of 63~mas based on a more recent \ac{vlbi} map by \cite{2004AA...426..463O}, and (ii) B3\,1133+432 (J1135+4258), which \cite{2002AA...389..115D} measure as 45~mas, but we adopt a size of 29~mas based on a more recent map by \cite{2007ApJ...658..203H}.

We classified five sources as bona fide \acp{cso} that show both small and large scale structure, which we interpret as a signature of different epochs of activity: 0108+388 (J0111+3906) \citep{1990AA...232...19B,2003PASA...20..118S}, B2\,0402+379 (J0405+3803)\citep{2004ApJ...602..123M}, NGC~3894 (J1148+5924) \citep{1998ApJ...498..619T}, J1247+6723 \citep{2003PASA...20...16M},  and PKS\,B1345+125 (J1347+1217) \citep{2005AA...443..891S}.
In these five cases the classification as a \ac{cso} is based on the small scale structure with size $<1$\,kpc, whereas the larger structure exceeds the size threshold by a factor of order 10--1000. In all five cases there was a large drop in surface brightness, or an actual gap in the brightness distribution, between the small- and large-scale structure.

\begin{figure}
    \centering
    \includegraphics[width=\columnwidth]{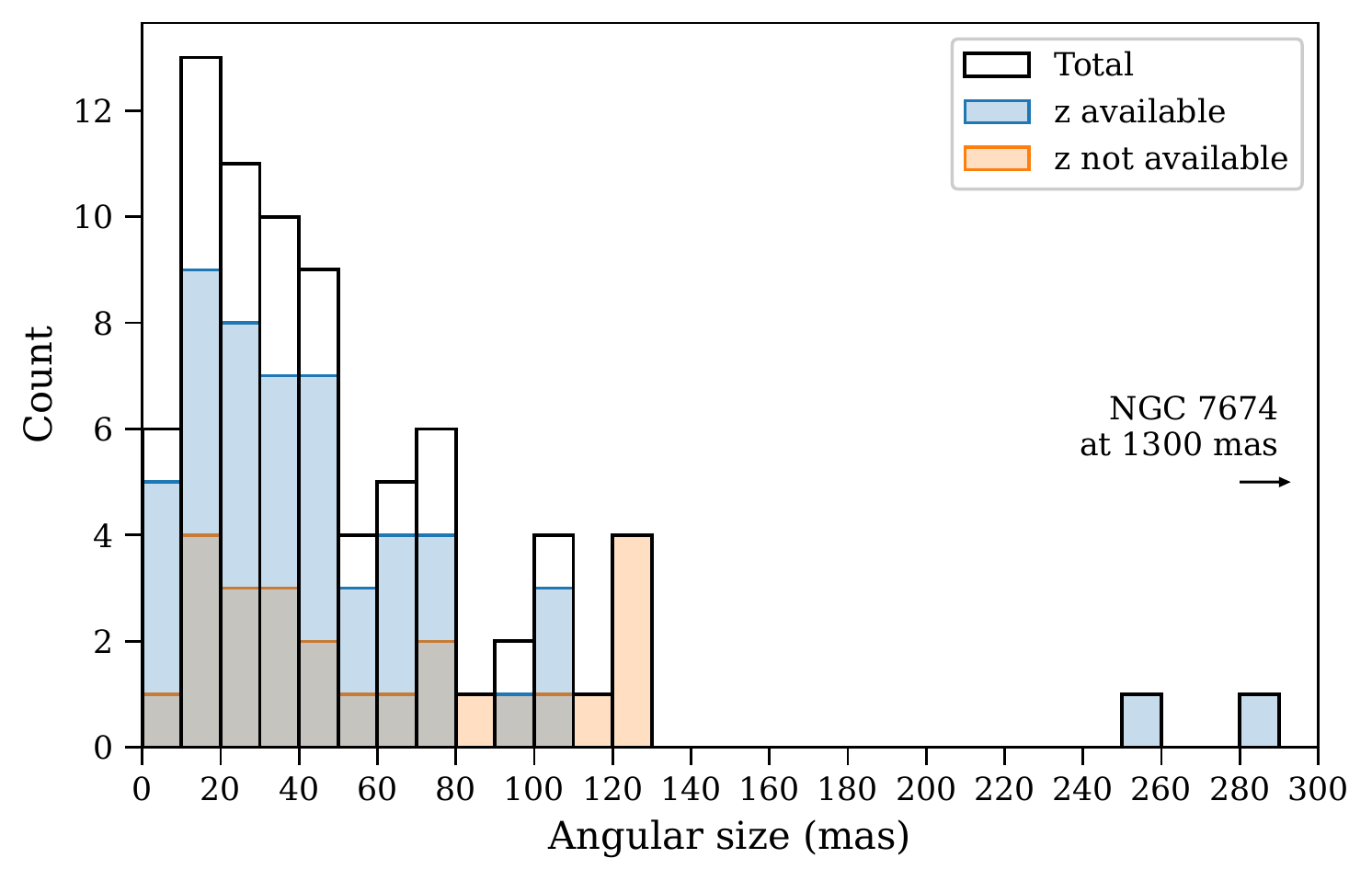}
    \caption{Angular size distribution of the bona fide \acp{cso}.}
    \label{fig:angsizedist}
\end{figure}

Our angular size measurements are listed in \cref{tab:csos} in Appendix~\ref{app:csocat} , and the  angular size distribution of the 79~\acp{cso} in our sample is shown in \cref{fig:angsizedist}. The two outlier CSOs at 250  mas and 280 mas are J1602+5243 and JJ1508+3423, respectively.  The distribution shows a dip below $\sim 10\,\mathrm{mas}$,  a peak from 10--50\,mas, and a steady decline in numbers at larger sizes. Both the strong cutoffs at small angular size and at large angular sizes are certain to be  strongly affected by the selection effects described in \cref{sec:bias-size} and hence cannot be used for statistical tests.

NGC~7674 (J2327+0846) is an outlier with a measured angular size of 1.3\,arcsec. It is a nearby Seyfert~2 galaxy with a redshift of 0.02892 \citep{2000AJ....120.1691N}. If it were located at the median bona fide \ac{cso} redshift ($z = 0.24$), it would have been detected as a \ac{cso} of $\sim 200\,\mathrm{mas}$~angular size, but probably would be too faint to image with \ac{vlbi}.

%------------------------------------------------------------------------------
\subsection{The Radio Spectra of CSOs}
\label{sec:csospecta}

The spectra of \acp{cso} are often characterized by a single peak in the $\sim 100\,\mathrm{MHz}$ to $\sim10\,\mathrm{GHz}$ range, although some show monotonically decreasing spectra.

The emission from CSOs is optically thin at frequencies above the peak and optically thick below the peak due either to free-free absorption (FFA) or synchrotron self-absorption (SSA). These must be due to the compactness of \acp{cso}.  In the case of free-free absorption, this would be a result of the  \ac{cso} emission regions being located in the dense central regions of their host galaxies.  In the case of synchrotron self-absorption it would arise because of the high surface brightness of the compact \ac{cso} emission regions \citep{1968ARAA...6..321S}.  In either case, the implied sizes of the  emission regions would range from a few tenths of a milliarcsecond for \acp{cso} with spectral peaks at a few GHz up to $\sim 10$\,mas for \acp{cso} with spectral peaks around 100\,MHz.   Since both FFA and SSA are expected to be stronger in more compact sources, it is to be expected that there is a relationship between the structure  and spectra of CSOs.
 This is an important part of the CSO story, not only because of the differences expected in CSO structures and spectra within a sample, but also  because we would expect individual CSO spectra to change as CSOs evolve and expand. We are engaged in a study of connections between the structure of CSOs in a complete sample and their spectra. While this is a very interesting subject, it is beyond the scope of this paper.

% FIGURE: spectral peak flux vs angular size
\begin{figure}
    \centering
    \includegraphics[width=\columnwidth]{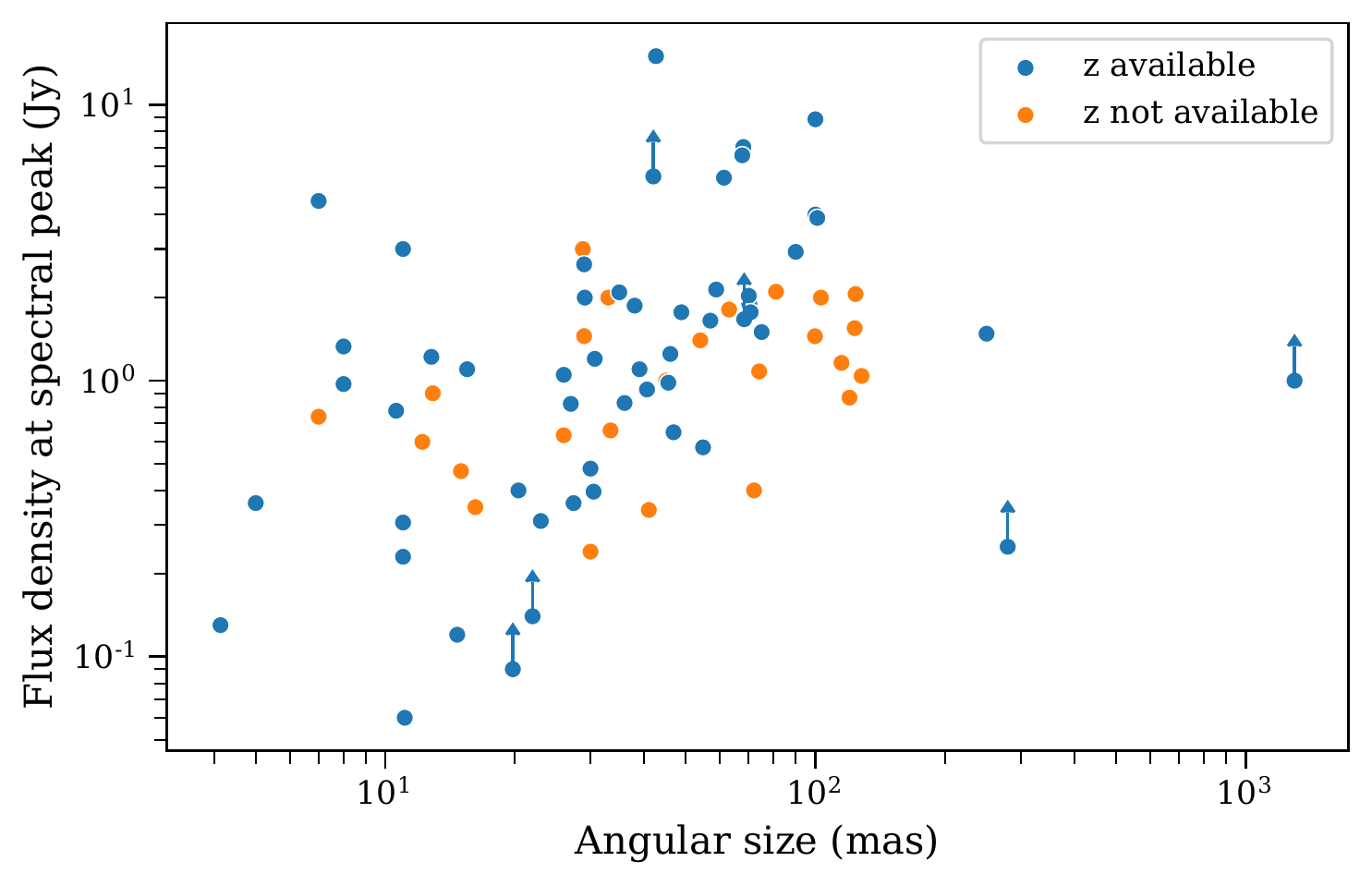}
    \caption{Flux density at radio spectral peak versus largest angular size for the bona fide \acp{cso}. The \acp{cso} plotted in blue have measured redshifts.}
    \label{fig:turnoverfluxvsangularsize}
\end{figure}

% FIGURE: turnover frequency vs angular size
\begin{figure}
    \centering
    \includegraphics[width=\columnwidth]{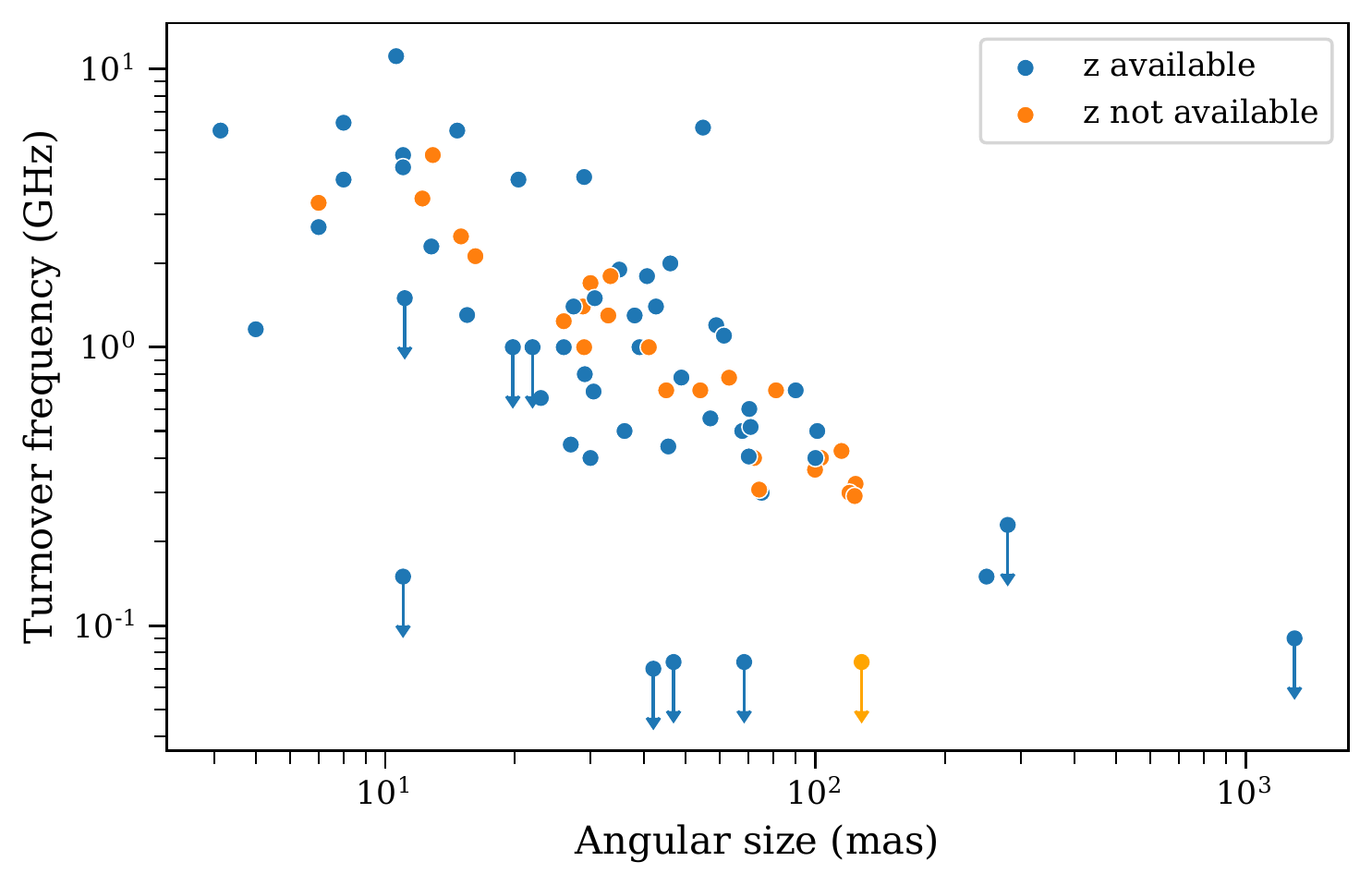}
    \caption{ Observed turnover frequency versus largest angular size for the bona fide \acp{cso}.}
    \label{fig:turnoverfreqvsangularsize}
\end{figure}

The flux densities at the spectral peak and the  angular sizes of our bona fide \acp{cso} are shown in \cref{fig:turnoverfluxvsangularsize}, and
in \cref{fig:turnoverfreqvsangularsize} we plot the turnover frequency ($\nu_m$) versus largest angular size. The values are listed in \cref{tab:csos} in Appendix~\ref{app:csocat} . The $\nu_m$ values were taken from the literature, or estimated by us based on multi-epoch radio spectra compiled on the \ac{rfc} website. The arrows indicate upper limits on $\nu_m$ due to a lack of low-frequency flux density measurements. 

\Cref{fig:turnoverfluxvsangularsize} shows a lower envelope on the spectral peak flux density that increases with angular size.  This is likely affected by the flux density limit bias described in \cref{sec:bias-flux}, as well as the angular size biases described in \cref{sec:bias-size}. It could also be the case that the peak flux density - size relationship seen here is influenced by evolution of the CSO. We return to this point in Paper 3.

Similarly, \cref{fig:turnoverfreqvsangularsize} shows a clear  decrease in turnover frequency with increasing  angular size. This is  likely caused by synchrotron self absorption and/or free-free absorption and/or CSO evolution \citep{1991ApJ...380...66O,1996ApJ...460..612R,2021AARv..29....3O}, and is also likely affected by the spectral index bias described in \cref{sec:bias-spectral}

\cref{fig:turnoverfreqvsangularsize} also illustrates another possible source of bias in the bona fide \ac{cso} size distribution, namely that many \ac{cso} surveys and compilations have targeted \ac{ps} sources, whereas \acp{cso} with sizes greater than 100~mas have $\nu_m$ below 500~MHz, as illustrated in the trend in \cref{fig:turnoverfreqvsangularsize}. 

Clearly, therefore, unless it can be shown that the effect under investigation will not be affected by these selection biases, in order to carry out any statistical tests on our sample of bona fide \acp{cso}, it is important to consider complete flux density limited samples in which no spectral index filtering has been applied.

% FIGURE: linear size distribution
\begin{figure}
    \centering
    \includegraphics[width=\columnwidth]{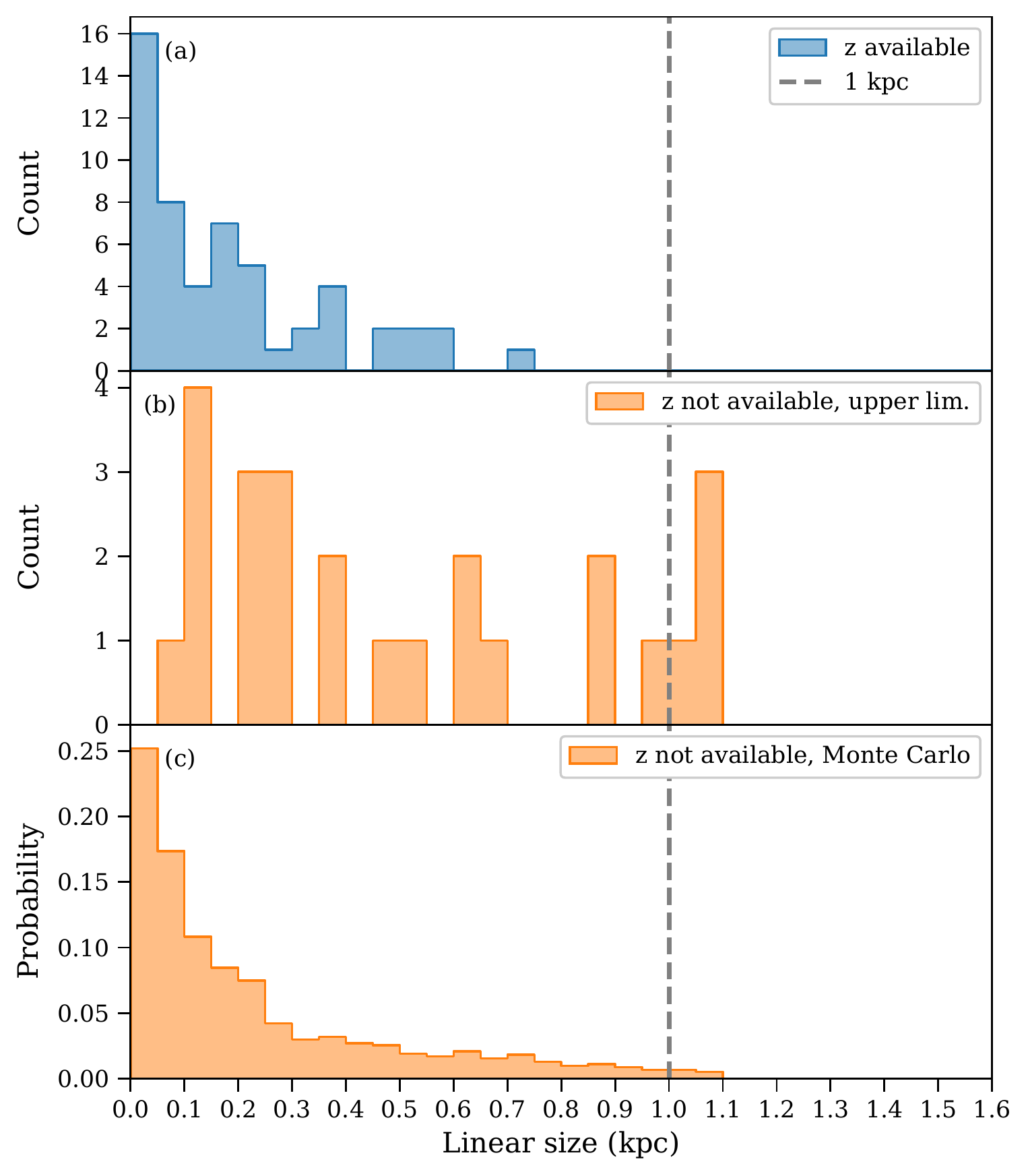}
    \caption{Linear size distribution of the bona fide \acp{cso}.
    Panel (a): histogram of the linear sizes of the 54~of~79~sources for which spectroscopic redshift estimates are available.
    Panel (b): histogram of the linear size upper limits for the 25~of~79~sources for which spectroscopic redshift estimates are not available. 
    Panel (c): histogram of the Monte Carlo simulated distribution of linear sizes for the 25~of~79~sources without spectroscopic redshift estimates, assuming these sources follow the same redshift distribution as the other \acp{cso}.}
    \label{fig:linsizedist}
\end{figure}

%------------------------------------------------------------------------------
\section{The Physical Sizes of CSO\lowercase{s}}
\label{sec:csos-linear-sizes}

We give the projected physical sizes for the 54~\acp{cso} for which we have spectroscopic redshifts  in \cref{tab:csos} in Appendix~\ref{app:csocat} , and show the  linear size distribution  in \cref{fig:linsizedist}~(a).

% without redshift - upper limits:
For the remaining 25~bona fide \acp{cso}  spectroscopic redshifts are not available and their location in the linear size distribution is unclear. Given the cosmology assumed throughout this paper, redshift~1.64 corresponds to the largest angular diameter distance. Assuming all \acp{cso} without redshift estimates are located at this redshift, we calculate the upper limits on the linear sizes, shown in \cref{fig:linsizedist}~(b). 
21~of these sources have angular sizes small enough that they will not exceed 1\,kpc in linear size regardless of their redshift.
The remaining four have angular sizes large enough that they would exceed 1\,kpc if they happen to lie in a specific redshift range:  B3\,1441+409 (J1443+4044,  $1.1 < z < 2.6$), B3\,2358+406 (J0000+4054, $1 < z < 2.8$), B3\,0233+434 (J0237+4342, $0.7 < z < 3.6)$, and J1928+6815 ($0.45 < z < 6.4$).

If we assume that the 25~\acp{cso} without spectroscopic redshifts follow the same redshift distribution as the 54~\acp{cso} for which redshift measurements are available, and randomly draw 10\,000~redshifts from the distribution shown in \cref{fig:redshiftdist} for each of the 25~sources, and take the mean of these, we get the distribution shown in \cref{fig:linsizedist}~(c).  There is a low probability that some or all of the four sources discussed above exceed 1\,kpc. The distribution drops gradually towards a linear size of 1\,kpc; this suggests that the linear size distribution of all 79~\acp{cso} would not differ significantly from the one shown in \cref{fig:linsizedist}~(a).

%------------------------------------------------------------------------------
\section{Conclusions}
\label{sec:conclusions}

The class of \acp{cso} was originally defined in order to enable detailed physical studies of the phenomenology of jetted-\acp{agn} without the confusion that arises in observing strongly relativistically boosted emission regions.  By adding two more selection criteria, based on variability and speed, we have laid the foundation for compiling a comprehensive catalog of \acp{cso} that are uncontaminated by \acp{agn} misidentified as \acp{cso}, thereby opening the way to the study of the phenomenology of \acp{cso}.

Through an extensive literature search we compiled a list of 3,175~candidate \acp{cso} and identified a sample of 79~bona fide \acp{cso},  including 15 newly-identified CSOs. In addition we identified 167~\ac{cso} A-class candidates that we are currently observing with the \ac{vlba}, in order to make a definitive decision on whether or not they are bona fide \acp{cso}.  The follow-up of the 1166~\ac{cso} B-class candidates will take much longer since there are so many of them.  

Our sample of 79~bona fide \acp{cso} includes  CSOs from three complete samples in  which we have identified all of the CSOs. If we include the two incomplete flat-spectrum limited samples  the fraction of \acp{cso}  ranges from 2.9\% to 9.4\%  The lower fractions occur in the incomplete flat-spectrum \ac{cjf} and \ac{vips} samples. These are  missing many steep-spectrum \acp{cso} due to their spectral selection filter.  Based on the complete \ac{pr}, \ac{cj1} and \ac{pw} samples, which have no spectral index  limits, the \ac{cso} fraction is $(6.8\pm 1.6)\% \rightarrow (8.5 \pm 1.8)\%$.  These 79~bona fide \acp{cso}, are suitable for some, but certainly not all, statistical studies. Any statistical studies should consider  the selection effects we have described.    We have initiated a program to expand the numbers of CSOs in complete samples by at least a factor 3.

In Appendix~\ref{app:csocat} we present a catalog of \acp{cso} which we anticipate will grow over time as new \acp{cso} are discovered. We hope that this effort, including our new criteria to eliminate highly beamed sources, will help to promote the study of the phenomenology of this distinctive class of jetted-\acp{agn}.

In two further papers, Paper~2 and Paper~3, we present compelling evidence that \acp{cso} comprise a distinct and separate class of jetted-\acp{agn}, and that most CSOs go through their whole  $\lesssim 5000$ yr lifespan, from early-, mid-, to late-life, as \acp{cso}, and we discuss the origin of \acp{cso}.  

\acp{cso} provide a unique window on relativistic jets and the central engines that drive them.  With the uncontaminated elucidation of their multi-wavelength phenomenology, CSOs  are poised to address, in completely new ways, the origins of relativistic jets and the basic properties of their central engines.

%------------------------------------------------------------------------------
\begin{acknowledgments}

  We thank the reviewer of this paper for many helpful suggestions that have clarified several important aspects of this work.
This research has made use of NASA’s Astrophysics Data System Bibliographic Services.
% NED:
This research has made use of the NASA/IPAC Extragalactic Database (NED) which is operated by the Jet Propulsion Laboratory, California Institute of Technology, under contract with the National Aeronautics and Space Administration.
% CATS:
The authors are grateful for the use of the CATS database of \citet{2005BSAO...58..118V}, of the Special Astrophysical Observatory. 
% OVRO:
This research has made use of data from the OVRO 40-m monitoring program (Richards, J. L. et al. 2011, ApJS, 194, 29), supported by private funding from the California Insitute of Technology and the Max Planck Institute for Radio Astronomy, and by NASA grants NNX08AW31G, NNX11A043G, and NNX14AQ89G and NSF grants AST-0808050, AST-1109911 and AST-1835400.
% MOJAVE:
This research has made use of data from the MOJAVE database that is maintained by the MOJAVE team \citep{2018ApJS..234...12L}. The MOJAVE program was supported by NASA-{\it Fermi} grant 80NSSC19K1579.
% SK
SK and KT acknowledge support from the European Research Council (ERC) under the European Unions Horizon 2020 research and innovation programme under grant agreement No.~771282.  
% KT
KT acknowledges support from the Foundation of Research and Technology - Hellas Synergy Grants Program through project POLAR, jointly implemented by the Institute of Astrophysics and the Institute of Computer Science.
%AS
AS acknowledges support from the NASA contract NAS8-03060 (Chandra X-ray Center).
%NRAO
This paper depended on a very large amount of VLBI data, almost all of which was taken with the Very Long Baseline Array. The National Radio Astronomy Observatory is a facility of the National Science Foundation operated under cooperative agreement by Associated Universities, Inc.

\end{acknowledgments}

\software{Astropy \citep{astropy:2013, astropy:2018, astropy:2022}, Astroquery \citep{2019AJ....157...98G}, Matplotlib \citep{Hunter:2007}, NumPy \citep{harris2020array}, pandas \citep{mckinney-proc-scipy-2010}, SciPy \citep{2020SciPy-NMeth}, seaborn \citep{Waskom2021}.}

\clearpage

%------------------------------------------------------------------------------
\appendix
%-------------

\section{Other important aspects of CSO\lowercase{s}}
\label{app:mwl}

Other important phenomenological aspects of \acp{cso},  include the properties of \acp{cso} in the  optical, infrared, X-ray, and $\gamma$-ray bands; the relationship of \acp{cso} to \acp{css} and \acp{ps} sources; and the cosmological evolution of \acp{cso}.
Many of these are beyond the scope of this study, but we include below brief summaries of the situation in other wavebands.

%------------------------------------------------------------------------------
\subsection{Optical properties of CSOs}
\label{app:mwl-optical}

The complete sample of PR has been observed at Palomar with the 200-inch telescope and very high quality spectra have been produced and reduced, including identification of all the emission lines, measurements of the equivalent widths, etc. \citep{1996ApJS..107..541L}, and the properties of some of the \ac{pr} \acp{cso} have been discussed in detail by R96.  We clearly need to carry out a study of all the \acp{cso} in the \ac{pr}~sample and to obtain similar quality spectra for the \acp{cso} in the \ac{cj1} and \ac{pw}~samples. A detailed discussion of the optical spectrum of 2352+495 (J2355+4950) is given in R96, and three other \acp{cso} are discussed in that paper.  We have undertaken a program to obtain similar quality optical spectra of all the \acp{cso} in the \ac{pr}, \ac{cj1}, and \ac{pw} samples.

%------------------------------------------------------------------------------
\subsection{Infrared properties of CSOs}
\label{app:mwl-infrared}

%Heckman et al 1994
\cite{1994ApJ...428...65H} compiled a large sample of radio galaxies and quasars from the IRAS database, which included compact \ac{ps}/\ac{css} sources. 
%Fanti et al  
The \cite{2000AA...358..499F} ISO~sample of \ac{ps}/\ac{css} sources included seven bona~fide \acp{cso} and two \ac{cso} candidates. These observations show no significant differences between the compact radio sources and large scale radio galaxies. However, \ac{cso} sources are definitely a distinct class, and not a subset of \ac{ps}/\ac{css} sources, as we show in Paper~2.

High quality {\it Spitzer} IRS~spectra for eight bona~fide \acp{cso} ($z<0.1$) have been presented by \cite{2010ApJ...713.1393W}.
%Willet et al. (2010). 
They show that the \ac{pah} emission lines are consistent with the presence of a dusty torus and a weak quasar-like nucleus. Additionally, a comparison with the \ac{agn}-dominated and starburst-dominated sample of galaxies indicates a mixture of both components present in these \acp{cso}. 

The \cite{2020ApJ...897..164K}
%Kosmaczewski et al. (2021) 
sample of 29~radio sources includes 22~bona~fide \acp{cso} and four \ac{cso} candidates. These authors studied the mid-infrared properties of the host galaxy and \ac{agn} component using WISE colors supplemented by the IRAS and Spitzer data. Their main conclusion was that the \ac{cso} host galaxies are mainly red ellipticals, some with distorted morphology. In the analysis of the WISE colors the \acp{cso} seem to be different from the evolved \ac{fr}\,II galaxies.

%------------------------------------------------------------------------------
\subsection{X-ray and gamma-ray emitting CSOs}
\label{app:mwl-highenergy}

There are only 26~bona fide \acp{cso} studied in X-rays to date. The \acp{cso} are X-ray faint and are not detected in the all sky X-ray surveys. The studies of their X-ray properties became possible with the Chandra X-ray Observatory ({\it Chandra}) and the XMM-Newton mission \citep{2009AN....330..264S,2016AN....337...52M}
%(see Siemiginowska 2009, Migliori 2016 for review) 
The first high quality X-ray spectrum of a \ac{cso}, OQ\,208 (J1407+2827), obtained by XMM-Newton \citep{2004AA...421..461G}, 
%(Guainazzi et al 2004). 
indicated that multiple emission and absorption components are present in the unresolved $\sim$ 3\,kpc central region of the host galaxy. A recent NuSTAR observation of this source gives the first broad energy coverage into the hard X-rays ($>10$\,keV) allowing for good constraints of the primary and scattered emission and the absorption by a high density porous structure \citep{2019ApJ...884..166S}.
%(Sobolewska et al 2019).

Small samples of \acp{cso} were observed with XMM-Newton
and {\it Chandra} \citep{2006AA...446...87G, 2006MNRAS.367..928V, 2008ApJ...684..811S, 2009AA...501...89T, 2016ApJ...823...57S},
% Vink et al 2006, Siemiginowska et al 2016) 
giving simple detections and the X-ray flux measurements required for
longer follow up observations. {\it Chandra} observations provide the highest angular resolution X-ray images, but they cannot resolve the X-ray emission on the scales of the double radio structures contained within $< 1\,\mathrm{kpc}$. However, X-rays detected on larger scales can probe the host galaxy environment or the presence of the structures potentially linked to the past activity. {\it Chandra} observations of PKS\,1718$-$649 (J1723-6500) show the X-rays diffuse emission extending out to about 2.5\,kpc distance from the nucleus and linked to a starburst activity in the central regions of the host galaxy. The X-ray emission originating in the central region of $<0.4$\,kpc could be associated with the accretion flow or compact radio lobes.

Three bona fide \acp{cso} have been detected in $\gamma$-rays with \textit{Fermi}-LAT: PKS~1718$-$649 \citep{2016ApJ...821L..31M}, TXS~0128+554 (J0131+5545) \citep{2020ApJ...899..141L} and NGC~3894 (J1148+5924) \citep{2020AA...635A.185P}. Given the radio morphology of CSOs, $\gamma$-rays are predicted to originate in the radio lobes
\citep{2008ApJ...680..911S,2021MNRAS.507.4564P}, and this process was successful in explaining the $\gamma$-rays detected in PKS~1718$-$649 
\citep{2022ApJ...941...52S}.
However, one interpretation of the radio structure of TXS~0128+554 is that the brightest component, which has a flat spectrum, is the core, and thus more typical of the radio-loud quasars, in which case the core could be the main source of the gamma-rays \citep{2020ApJ...899..141L}. In Paper 3 we present an alternative interpretation  of the radio structure of TXS~0128+554, in which the brightest component is not the core, and so the core in this alternative interpretation is less likely to be the origin of the gamma-rays. Recent modeling of the PKS~1718$-$649 spectral energy distribution 
\cite{2021MNRAS.507.4564P}
performed analysis of the \textit{Fermi}-LAT data for a large sample of \ac{ps}, \ac{css} and \ac{cso} galaxies and quasars. They concluded that the $\gamma$-ray radiation from galaxies is quite faint and only low redshift sources could be detected if the radiation is non-beamed.

%-----------------------------------------------------------------
\section{Table  of Bona Fide CSO\lowercase{s}}
\label{app:csocat}

\Cref{tab:csos} lists the 79~bona fide \acp{cso} and their properties. 
 In the following we list the references corresponding to all index numbers used in \cref{tab:csos}. The references are grouped by the type of value they refer to. 

 Redshift references:
$^{1}$ \citet{2002AJ....124..662Z}, $^{2}$ \citet{1998ApJ...494..175C}, $^{3}$ \citet{2007AA...468L..71G}, $^{4}$ \citet{2012ApJS..199...26H}, $^{5}$ \citet{2009AA...496L...9M}, $^{6}$ \citet{1986AJ.....91..494L}, $^{7}$ \citet{2015MNRAS.452.4153A}, $^{8}$ \citet{1998MNRAS.295..946R}, $^{9}$ \citet{2008ApJS..175..297A}, $^{10}$ \citet{2017ApJS..233...25A}, $^{11}$ \citet{2020ApJS..249....3A}, $^{12}$ \citet{1996ApJS..107..541L}, $^{13}$ \citet{2000ApJ...534..104P}, $^{14}$ \citet{1995AAS..114..259D}, $^{15}$ \citet{2002MNRAS.329..877C}, $^{16}$ \citet{2011ApJS..193...29A}, $^{17}$ \citet{1997ApJS..112..391H}, $^{18}$ \citet{1992ApJS...81...83H}, $^{19}$ \citet{1995AJ....109...14O}, $^{20}$ \citet{1996AJ....111.1013V}, $^{21}$ \citet{2008MNRAS.387..639H}, $^{22}$ \citet{2006ApJS..162...38A}, $^{23}$ \citet{2010AJ....139.2360S}, $^{24}$ \citet{1998ApJ...494...47F}, $^{25}$ \citet{2016MNRAS.459..820T}, $^{26}$ \citet{1993ApJ...409..170M}, $^{27}$ \citet{2001ARep...45...79C}, $^{28}$ \citet{2007AA...463...97L}, $^{29}$ \citet{2014MNRAS.440..696A}, $^{30}$ \citet{2007AA...464..879D}, $^{31}$ \citet{2008AA...484..119B}, $^{32}$ \citet{1997MNRAS.290..380H}, $^{33}$ \citet{1987MNRAS.225..761F}, $^{34}$ \citet{1993AAS..100..395S}, $^{35}$ \citet{1984ApJ...279..112B}, $^{36}$ \citet{2008ApJS..175...97H}, $^{37}$ \citet{2000AJ....120.1691N}.

 All angular sizes have been measured newly in this work for the CSOs listed in \cref{tab:csos}. The following notes give each reference from which the map for the size measurement was selected and highlights the selected map frequency:
$^{1}$ 1.67\,GHz, \citet{2002AA...389..115D}, $^{2}$ 4.975\,GHz, \citet{2000ApJ...534...90P}, $^{3}$ 2.3\,GHz, \citet{2011AA...535A..24S}, $^{4}$ 15\,GHz, MOJAVE - the MOJAVE stacked 15\,GHz image shows the largest angular size compared to lower frequency maps, $^{5}$ 1.3\,GHz, \citet{2003AA...399..889G}, $^{6}$ 15\,GHz, \citet{2020ApJ...899..141L}, $^{7}$ 8.4\,GHz, \citet{2012ApJS..198....5A}, $^{8}$ 1.6\,GHz, \citet{2002AA...389..115D}, $^{9}$ 8.4\,GHz, \citet{2005ApJ...622..136G}, $^{10}$ 5\,GHz, \citet{1998MNRAS.299.1159A}, $^{11}$ 5\,GHz, \citet{1998MNRAS.299.1159A}, $^{12}$ 5\,GHz, \citet{2004ApJ...602..123M}, $^{13}$ 5\,GHz, \citet{2003ApJ...597..157T}, $^{14}$ 8.5\,GHz, \citet{2005ApJ...622..136G}, $^{15}$ 1.4\,GHz, \citet{2000MNRAS.319..429S}, $^{16}$ 5\,GHz, \citet{2004AA...426..463O}, $^{17}$ 15\,GHz, MOJAVE, $^{18}$ 5\,GHz, \citet{2003ApJ...597..157T}, $^{19}$ 5\,GHz, \citet{2016MNRAS.459..820T}, $^{20}$ 5\,GHz, \citet{2016MNRAS.459..820T}, $^{21}$ 1.6\,GHz, \citet{2002AA...389..115D}, $^{22}$ 5\,GHz, \citet{2007ApJ...658..203H}, $^{23}$ 1.4\,GHz, \citet{2016MNRAS.462.2819B}, $^{24}$ 5\,GHz, \citet{2021MNRAS.506.1609C}, $^{25}$ 5\,GHz, \citet{2021MNRAS.506.1609C}, $^{26}$ 5\,GHz, \citet{2007ApJ...658..203H}, $^{27}$ 1.6\,GHz, \citet{2002AA...389..115D}, $^{28}$ 5\,GHz, \citet{2021MNRAS.506.1609C}, $^{29}$ \,GHz, \citet{2006MNRAS.368.1411A}, $^{30}$ 5\,GHz, \citet{2016MNRAS.459..820T}, $^{31}$ 2.3\,GHz, RFC (2018-03-26), $^{32}$ 2.3\,GHz, RFC 2017-01-16, $^{33}$ 5\,GHz, \citet{2007ApJ...658..203H}, $^{34}$ 2.3\,GHz, RFC 2017-08-05, $^{35}$ 5\,GHz, \citet{2016MNRAS.459..820T}, $^{36}$ 5\,GHz, \citet{2016MNRAS.459..820T}, $^{37}$ 5\,GHz, \citet{2016MNRAS.459..820T}, $^{38}$ 4.99\,GHz, \citet{2021MNRAS.506.1609C}, $^{39}$ 2.3\,GHz, RFC 2014-5-31, $^{40}$ 5\,GHz, \citet{2016MNRAS.459..820T}, $^{41}$ 2.3\,GHz, RFC 2017-06-10, $^{42}$ 1.4\,GHz, \citet{2002AA...389..115D}, $^{43}$ 2.3\,GHz, \citet{2011AA...535A..24S}, $^{44}$ 5\,GHz, \citet{2016MNRAS.459..820T}, $^{45}$ 5\,GHz, \citet{2016MNRAS.459..820T}, $^{46}$ 5\,GHz, \citet{2016MNRAS.459..820T}, $^{47}$ 5\,GHz, \citet{2016MNRAS.459..820T}, $^{48}$ 1.6\,GHz, \citet{2014MNRAS.438..463O}, $^{49}$ 1.66\,GHz, \citet{2002AA...385..768X}, $^{50}$ 2\,GHz, RFC 2017-06-15, $^{51}$ 8.4\,GHz, \citet{2003ChJAA...3..505W}, $^{52}$ 5\,GHz, \citet{2016AJ....151...74Y}, $^{53}$ 5\,GHz, \citet{2007ApJ...658..203H}, $^{54}$ 1.6\,GHz, \citet{1995AA...295...27D}, $^{55}$ 1.6\,GHz, \citet{2002AA...389..115D}, $^{56}$ 5\,GHz, \citet{2016MNRAS.459..820T}, $^{57}$ 1.6\,GHz, \citet{2002AA...389..115D}, $^{58}$ 5\,GHz, \citet{2010MNRAS.408.2261K}, $^{59}$ 5\,GHz, \citet{2006AA...454..729X}, $^{60}$ 5\,GHz, \citet{2016MNRAS.459..820T}, $^{61}$ 1.6\,GHz, \citet{2009AA...498..641D}, $^{62}$ 2.2\,GHz, \citet{2006ApJ...648..148N}, $^{63}$ 5\,GHz, \citet{2016MNRAS.459..820T}, $^{64}$ 4.9\,GHz, \citet{2002ApJS..141..311T}, $^{65}$ 2.3\,GHz, \citet{2011AA...535A..24S}, $^{66}$ 8.4\,GHz, \citet{2002AA...385..768X}, $^{67}$ 8.5\,GHz, \citet{2005ApJ...622..136G}, $^{68}$ 4.8\,GHz, \citet{2007ApJ...661...78G}, $^{69}$ 1.65\,GHz, \citet{2007AA...470...97L}, $^{70}$ 4.8\,GHz, \citet{2007ApJ...661...78G}, $^{71}$ 5\,GHz, \citet{1998MNRAS.299.1159A}, $^{72}$ 8.4\,GHz, \citet{2004AJ....127.1977O}, $^{73}$ 1.6\,GHz, \citet{2003PASA...20...19M}, $^{74}$ 8\,GHz, \citet{1999ApJ...521..103P}, $^{75}$ 2.3\,GHz, \citet{2011AA...535A..24S}, $^{76}$ 8.42\,GHz, \citet{2006AA...450..959O}, $^{77}$ 1.4\,GHz, \citet{2003ApJ...597..809M}, $^{78}$ 2.3\,GHz, \citet{2011AA...535A..24S}, $^{79}$ 0.6\,GHz, \citet{1996ApJ...460..612R}.

Turnover frequency and flux density references:
$^{1}$ Estimated by the authors from RFC spectrum, $^{2}$ Estimated by authors from NED spectrum, $^{3}$ \citet{2016MNRAS.458.3786J}, $^{4}$ \citet{2019AstBu..74..348S}, $^{5}$ \citet{2012ApJ...760...77A}, $^{6}$ \citet{2020ApJ...899..141L}, $^{7}$ \citet{1999AAS..135..273M}, $^{8}$ \citet{2017ApJ...836..174C}, $^{9}$ \citet{2007AA...463...97L}, $^{10}$ \citet{2006AA...454..729X}, $^{11}$ \citet{2004MNRAS.348..227S}, $^{12}$ \citet{2007ApJ...661...78G}.

 X-ray detection references:
$^{1}$ \citet{2006AA...446...87G}, $^{2}$ \citet{2019ApJ...871...71S}, $^{3}$ \citet{2006MNRAS.367..928V}, $^{4}$ \citet{2009ApJ...690..644G}, $^{5}$ \citet{2016ApJ...823...57S}, $^{6}$ \citet{2020ApJ...899..141L}, $^{7}$ \citet{2020ApJ...899..141L}, $^{8}$ \citet{2014ApJ...780..149R}, $^{9}$ \citet{2017ApJ...835..223S}, $^{10}$ \citet{2010AA...517A..33Y}, $^{11}$ \citet{2009AA...501...89T}, $^{12}$ \citet{2008ApJ...684..811S}, $^{13}$ \citet{2009AN....330..264S}, $^{14}$ \citet{2009AA...501...89T}, $^{15}$ \citet{2004AA...421..461G}, $^{16}$ \citet{2019ApJ...884..166S}, $^{17}$ \citet{2010ApJS..189...37E}, $^{18}$ \citet{2018AA...612L...4B}.

 $\gamma$-ray detection references:
$^{1}$ \citet{2020ApJ...899..141L}, $^{2}$ \citet{2020arXiv201008406L}, $^{3}$ \citet{2021MNRAS.507.4564P}, $^{4}$ \citet{2022ApJ...927..221G}, $^{5}$ \citet{2020AA...635A.185P}, $^{6}$ \citet{2016ApJ...821L..31M}, $^{7}$ \citet{2020ApJ...892..105A}.

 \ac{cso} and \ac{cso} candidate references:
$^{1}$ \citet{2000ApJ...534...90P}, $^{2}$ \citet{2002AA...389..115D}, $^{3}$ \citet{2005ApJ...622..136G}, $^{4}$ \citet{2006MNRAS.368.1411A}, $^{5}$ \citet{2007ApJ...661...78G}, $^{6}$ \citet{2009ARep...53..319T}, $^{7}$ \citet{2014ApJ...780..178M}, $^{8}$ \citet{1996ApJ...460..612R}, $^{9}$ \citet{1996ApJ...463...95T}, $^{10}$ \citet{1996PhDT........44A}, $^{11}$ \citet{1997hsra.book..153C}, $^{12}$ \citet{1998MNRAS.299.1159A}, $^{13}$ \citet{2005ApJ...618..635G}, $^{14}$ \citet{2009AA...505..509L}, $^{15}$ \citet{2010ApJ...713.1393W}, $^{16}$ \citet{2020ApJ...899..141L}, $^{17}$ \citet{2012ApJS..198....5A}, $^{18}$ \citet{2007AA...461..923O}, $^{19}$ \citet{2004ApJ...602..123M}, $^{20}$ \citet{2003ApJ...597..157T}, $^{21}$ \citet{2000MNRAS.319..429S}, $^{22}$ \citet{2016MNRAS.459..820T}, $^{23}$ \citet{2007ApJ...658..203H}, $^{24}$ \citet{2005ApJS..159...27T}, $^{25}$ \citet{2016MNRAS.462.2819B}, $^{26}$ \citet{2011AA...528A.110F}, $^{27}$ \citet{1998MNRAS.297..559B}, $^{28}$ \citet{2005AA...434..123X}, $^{29}$ \citet{2006AA...454..729X}, $^{30}$ \citet{2020AA...635A.185P}, $^{31}$ \citet{2014MNRAS.438..463O}, $^{32}$ \citet{2002AA...385..768X}, $^{33}$ \citet{2003PASA...20..118S}, $^{34}$ \citet{2016AJ....151...74Y}, $^{35}$ \citet{2016AN....337..130J}, $^{36}$ \citet{2006ApJ...648..148N}, $^{37}$ \citet{2019AA...627A.148A}, $^{38}$ \citet{1999AAS..134..309S}, $^{39}$ \citet{2006AA...450..959O}, $^{40}$ \citet{1999ApJ...521..103P}, $^{41}$ \citet{2001ApJ...554L.147P}, $^{42}$ \citet{2009ApJ...698.1282T}, $^{43}$ \citet{1999NewAR..43..681T}, $^{44}$ \citet{2000AA...360..887T}, $^{45}$ \citet{2010AJ....139...17A}.

\begin{longrotatetable}
\begin{deluxetable*}{lllllrrlllp{2cm}}
    %\rotate
    \tablecaption{%
        Bona fide \acp{cso}. Columns show (1) the J2000 name, (2) a common name, (3,~4) J2000 right ascension and declination, (5) redshift, (6,~7) angular and linear size, (8,~9) turnover frequency and flux density, (10) whether the source is X-ray and/or $\gamma$-ray detected, and (11) references that discussed the source as \ac{cso} or \ac{cso} candidate. All values are index with reference numbers. All references corresponding to the index numbers are listed in Appendix~\ref{app:csocat}.
        \label{tab:csos}
        }
    \tablehead{%
        \colhead{J2000 name} & \colhead{Common name} & \colhead{R.A.} & \colhead{Dec.} & \colhead{Redshift} & \colhead{Ang.} & \colhead{Lin.} & \colhead{Turnover} & \colhead{Turnover} & \colhead{X- or} & \colhead{CSO reference} \\
         & & & & & \colhead{size} & \colhead{size} & \colhead{freq.} & \colhead{flux dens.} & \colhead{$\gamma$-ray} & \colhead{\phantom{x}} \\
         & & & & & \colhead{(mas)} & \colhead{(kpc)} & \colhead{(GHz)} & \colhead{(Jy)} & & \colhead{\phantom{x}}
         } 
    \startdata
        J0000+4054 & B3~2358+406 & 00:00:53.08 & +40:54:01.81 &  & 124.0$^{*,1}$ & $^\ddagger$ & $0.323^{3}$ & $2.06^{3}$ &  & 1, 2, 3, 4, 5, 6 \\
        J0003+4807 & JVAS~J0003+4807 & 00:03:46.04 & +48:07:04.14 &  & 16.2$^{*,2}$ & 0.139$^*$ & $2.123^{3}$ & $0.348^{3}$ &  & 1, 3, 4 \\
        J0029+3456 & B2~0026+34 & 00:29:14.24 & +34:56:32.25 & 0.517$^{1}$ & 29.1$^{*,3}$ & 0.180$^*$ & $0.8^{4}$ & $2.0^{4}$ & X$^\mathrm{1,2}$ & 4, 7 \\
        J0111+3906 & 0108+388 & 01:11:37.32 & +39:06:28.10 & 0.66847$^{2}$ & 8.0$^{*,4,m}$ & 0.056$^*$ & $4.0^{3}$ & $1.33^{3}$ & X$^\mathrm{2,3,4,5}$ & 4, 8, 9 \\
        J0119+3210 & B2~0116+31 & 01:19:35.00 & +32:10:50.06 & 0.0602$^{3}$ & 100.0$^{*,5}$ & 0.115$^*$ & $0.4^{5}$ & $4.0^{5}$ & X$^\mathrm{5}$ & 10, 11, 12, 13, 14, 15, 4 \\
        J0131+5545 & TXS~0128+554 & 01:31:13.82 & +55:45:12.98 & 0.03649$^{4}$ & 23.0$^{*,6}$ & 0.016$^*$ & $0.657^{6}$ & $0.31^{6}$ & $\gamma^\mathrm{1,2,3}$ & 16 \\
        J0132+5620 & JVAS~J0132+5620 & 01:32:20.45 & +56:20:40.37 &  & 12.2$^{*,7}$ & 0.104$^*$ & $3.42^{7}$ & $0.6^{7}$ &  & 1, 17, 3 \\
        J0150+4017 & B3~0147+400 & 01:50:19.61 & +40:17:30.02 &  & 103.0$^{*,8}$ & 0.882$^*$ & $0.4^{*,1}$ & $2.0^{*,1}$ &  & 18, 4 \\
        J0204+0903 & JVAS~J0204+0903 & 02:04:34.76 & +09:03:49.26 &  & 33.0$^{*,9}$ & 0.282$^*$ & $1.3^{4}$ & $2.0^{4}$ &  & 1, 3, 4 \\
        J0237+4342 & B3~0233+434 & 02:37:01.21 & +43:42:04.18 &  & 120.0$^{*,10}$ & $^\ddagger$ & $0.3^{*,1}$ & $0.868^{*,1}$ &  & 10, 12, 4 \\
        J0402+8241 & JVAS~J0402+8241 & 04:02:12.68 & +82:41:35.13 &  & 72.0$^{*,11}$ & 0.616$^*$ & $0.4^{*,2}$ & $0.4^{*,2}$ &  & 12, 4 \\
        J0405+3803 & B3~0402+379 & 04:05:49.26 & +38:03:32.24 & 0.05505$^{5}$ & 42.0$^{*,12,m}$ & 0.044$^*$ & $<0.07^{*,1}$ & $>5.5^{*,1}$ & X$^\mathrm{8}$ & 15, 19 \\
        J0425-1612 & PKS~0423-163 & 04:25:53.57 & -16:12:40.23 &  & 99.8$^{*,13}$ & 0.854$^*$ & $0.363^{8}$ & $1.449^{8}$ &  & 20, 4 \\
        J0427+4133 & B3~0424+414 & 04:27:46.05 & +41:33:01.10 &  & 7.0$^{*,14}$ & 0.060$^*$ & $3.3^{4}$ & $0.74^{4}$ &  & 1, 3, 4 \\
        J0440+6157 & GB6~J0440+6158 & 04:40:46.90 & +61:57:58.57 &  & 30.0$^{*,15}$ & 0.257$^*$ & $1.7^{4}$ & $0.24^{4}$ &  & 21 \\
        J0706+4647 & B3~0703+468 & 07:06:48.07 & +46:47:56.45 &  & 63.0$^{*,16}$ & 0.539$^*$ & $0.777^{3}$ & $1.81^{3}$ &  & 2, 4 \\
        J0713+4349 & B3~0710+439 & 07:13:38.16 & +43:49:17.21 & 0.518$^{6}$ & 35.0$^{*,17}$ & 0.217$^*$ & $1.9^{3}$ & $2.09^{3}$ & X$^\mathrm{2,3,5}$ & 4, 8, 9 \\
        J0735-1735 & PKS~0733-17 & 07:35:45.81 & -17:35:48.50 &  & 28.8$^{*,18}$ & 0.246$^*$ & $1.4^{4}$ & $3.0^{4}$ &  & 20 \\
        J0741+2706 & B2~0738+27 & 07:41:25.73 & +27:06:45.42 & 0.772139$^{7}$ & 26.0$^{*,19}$ & 0.193$^*$ & $1.0^{*,1}$ & $1.05^{*,1}$ &  & 22 \\
        J0754+5324 & JVAS~J0754+5324 & 07:54:15.22 & +53:24:56.45 &  & 26.0$^{*,20}$ & 0.223$^*$ & $1.24^{3}$ & $0.634^{3}$ &  & 1, 22, 23, 3, 4 \\
        J0825+3919 & B3~0822+394 & 08:25:23.68 & +39:19:45.76 & 1.21$^{8}$ & 70.7$^{*,21}$ & 0.591$^*$ & $0.517^{3}$ & $1.77^{3}$ & $\gamma^\mathrm{4}$ & 2 \\
        J0832+1832$^\dagger$ & PKS~0829+18 & 08:32:16.04 & +18:32:12.12 & 0.154$^{9}$ & 30.7$^{*,22}$ & 0.081$^*$ & $1.5^{*,1}$ & $1.2^{*,1}$ &  & 22, 23 \\
        J0855+5751 & JVAS~J0855+5751 & 08:55:21.36 & +57:51:44.09 & 0.025998$^{10}$ & 75.0$^{*,23}$ & 0.039$^*$ & $0.3^{*,1}$ & $1.5^{*,1}$ &  & 22, 23, 24, 25 \\
        J0906+4124$^\dagger$ & GB6~J0906+4124 & 09:06:52.80 & +41:24:30.00 & 0.0273577$^{11}$ & 11.1$^{*,24}$ & 0.006$^*$ & $<1.5^{*,2}$ & $0.06^{*,2}$ &  & \phantom{x} \\
        J0909+1928$^\dagger$ & MRK~1226 & 09:09:37.44 & +19:28:08.30 & 0.027843$^{11}$ & 14.7$^{*,25}$ & 0.008$^*$ & $6.0^{*,2}$ & $0.12^{*,2}$ &  & \phantom{x} \\
        J0943+1702 & JVAS~J0943+1702 & 09:43:17.23 & +17:02:18.97 & 1.601115$^{11}$ & 20.4$^{*,26}$ & 0.175$^*$ & $4.0^{*,1}$ & $0.4^{*,1}$ &  & 22, 23 \\
        J1011+4204 & B3~1008+423 & 10:11:54.18 & +42:04:33.38 &  & 115.0$^{*,27}$ & 0.984$^*$ & $0.424^{3}$ & $1.16^{3}$ &  & 2 \\
        J1025+1022$^\dagger$ & NVSS~J102544+102231 & 10:25:44.20 & +10:22:30.00 & 0.045805$^{4}$ & 19.8$^{*,28}$ & 0.018$^*$ & $<1.0^{*,2}$ & $>0.09^{*,2}$ &  & \phantom{x} \\
        J1035+5628 & JVAS~J1035+5628 & 10:35:07.04 & +56:28:46.79 & 0.46$^{12}$ & 38.0$^{*,29}$ & 0.221$^*$ & $1.3^{3}$ & $1.87^{3}$ & X$^\mathrm{2,3,5}$ & 22, 23, 4, 6, 8, 9 \\
        J1042+2949 & B2~1039+30B & 10:42:36.51 & +29:49:45.15 &  & 45.0$^{*,30}$ & 0.385$^*$ & $0.7^{*,1}$ & $1.0^{*,1}$ &  & 22, 23 \\
        J1111+1955 & PKS~1108+201 & 11:11:20.07 & +19:55:36.01 & 0.299$^{13}$ & 15.5$^{*,31}$ & 0.068$^*$ & $1.305^{3}$ & $1.1^{3}$ &  & 1, 22, 23, 26, 3, 4, 6 \\
        J1120+1420 & PKS~1117+146 & 11:20:27.81 & +14:20:54.97 & 0.362$^{14}$ & 101.0$^{*,32}$ & 0.507$^*$ & $0.5^{3}$ & $3.89^{3}$ & X$^\mathrm{4}$ & 26, 27, 4 \\
        J1135+4258 & B3~1133+432 & 11:35:55.99 & +42:58:44.65 &  & 29.0$^{*,33}$ & 0.248$^*$ & $1.0^{9}$ & $1.45^{9}$ &  & 2, 22, 23, 28, 29 \\
        J1148+5924 & NGC~3894 & 11:48:50.36 & +59:24:56.36 & 0.01075$^{15}$ & 54.8$^{*,34,m}$ & 0.012$^*$ & $6.149^{3}$ & $0.573^{3}$ & $\gamma^\mathrm{2,3,5}$ & 15, 22, 23, 30 \\
        J1158+2450 & PKS~1155+251 & 11:58:25.79 & +24:50:18.00 & 0.203$^{16}$ & 46.0$^{*,35}$ & 0.152$^*$ & $2.0^{*,1}$ & $1.25^{*,1}$ &  & 22 \\
        J1159+5820 & VERA~J1159+5820 & 11:59:48.77 & +58:20:20.31 & 1.27997$^{11}$ & 70.2$^{*,36}$ & 0.591$^*$ & $0.6^{*,1}$ & $1.9^{*,1}$ &  & 22, 23 \\
        J1204+5202 & GB6~J1204+5202 & 12:04:18.61 & +52:02:17.62 &  & 54.0$^{*,37}$ & 0.462$^*$ & $0.7^{*,1}$ & $1.4^{*,1}$ &  & 22, 23 \\
        J1205+2031$^\dagger$ & NGC~4093 & 12:05:51.50 & +20:31:19.00 & 0.02378857$^{11}$ & 22.0$^{*,38}$ & 0.010$^*$ & $<1.0^{*,2}$ & $>0.14^{*,2}$ &  & \phantom{x} \\
        J1220+2916 & NGC~4278 & 12:20:06.82 & +29:16:50.72 & 0.002$^{17}$ & 46.8$^{*,39}$ & 0.002$^*$ & $<0.074^{*,1}$ & $0.65^{*,1}$ & X$^\mathrm{10,2}$ & 22, 23 \\
        J1227+3635 & B21225+36 & 12:27:58.72 & +36:35:11.82 & 1.975$^{18}$ & 58.8$^{*,40}$ & 0.499$^*$ & $1.2^{9}$ & $2.14^{9}$ &  & 22, 23, 26 \\
        J1234+4753 & JVAS~J1234+4753 & 12:34:13.33 & +47:53:51.24 & 0.373082$^{11}$ & 27.4$^{*,41}$ & 0.140$^*$ & $1.4^{*,1}$ & $0.36^{*,1}$ & X$^\mathrm{2,4}$ & 22, 23 \\
        J1244+4048 & B3~1242+410 & 12:44:49.19 & +40:48:06.15 & 0.813586$^{11}$ & 70.0$^{*,42}$ & 0.529$^*$ & $0.405^{3}$ & $2.03^{3}$ &  & 2, 22, 23, 26 \\
        J1247+6723$^\dagger$ & JVAS~J1247+6723 & 12:47:33.33 & +67:23:16.45 & 0.107219$^{11}$ & 5.0$^{*,43,m}$ & 0.010$^*$ & $1.16^{7}$ & $0.36^{7}$ & X$^\mathrm{5}$ & \phantom{x} \\
        J1254+1856 & CRATES~J1254+1856 & 12:54:33.27 & +18:56:01.93 & 0.1145$^{19}$ & 4.14$^{*,44}$ & 0.008$^*$ & $6.0^{*,2}$ & $0.13^{*,2}$ &  & 22, 23 \\
        J1311+1658 & JVAS~J1311+1658 & 13:11:23.82 & +16:58:44.22 & 0.081408$^{10}$ & 27.0$^{*,45}$ & 0.041$^*$ & $0.447^{8}$ & $0.824^{8}$ &  & 1, 22, 23, 3 \\
        J1313+5458$^\dagger$ & JVAS~J1313+5458 & 13:13:37.85 & +54:58:23.91 & 0.613$^{20}$ & 57.0$^{*,46}$ & 0.384$^*$ & $0.555^{7}$ & $1.65^{7}$ &  & 22, 23 \\
        J1326+3154 & DA~344 & 13:26:16.51 & +31:54:09.52 & 0.36801$^{21}$ & 68.0$^{*,47}$ & 0.345$^*$ & $0.5^{3}$ & $7.03^{3}$ & X$^\mathrm{11,2,4}$ & 22, 23, 26, 4 \\
        J1335+5844 & JVAS~J1335+5844 & 13:35:25.93 & +58:44:00.29 &  & 12.9$^{*,48}$ & 0.110$^*$ & $4.9^{9}$ & $0.9^{9}$ &  & 17, 22, 23, 29, 31 \\
        J1347+1217 & PKS~B1345+125 & 13:47:33.36 & +12:17:24.24 & 0.121$^{22}$ & 100.0$^{*,49,m}$ & 0.215$^*$ & $0.4^{3}$ & $8.86^{3}$ & X$^\mathrm{12,13}$ & 15, 32, 33, 34 \\
        J1400+6210 & 1358+625 & 14:00:28.65 & +62:10:38.59 & 0.431$^{6}$ & 67.6$^{*,50}$ & 0.378$^*$ & $0.5^{3}$ & $6.56^{3}$ & X$^\mathrm{14,2,3,4}$ & 26, 4, 6, 8, 9 \\
        J1407+2827 & OQ~+208 & 14:07:00.40 & +28:27:14.69 & 0.077$^{23}$ & 11.0$^{*,51}$ & 0.016$^*$ & $4.9^{10}$ & $3.0^{10}$ & X$^\mathrm{15,16,2,5}$ & 15, 26, 29, 32, 35, 4 \\
        J1413+1509 & JVAS~J1413+1509 & 14:13:41.66 & +15:09:39.51 &  & 15.0$^{*,52}$ & 0.128$^*$ & $2.5^{*,2}$ & $0.47^{*,2}$ &  & 22, 23, 34 \\
        J1414+4554 & B3~1412+461 & 14:14:14.85 & +45:54:48.73 & 0.186$^{24}$ & 30.5$^{*,53}$ & 0.094$^*$ & $0.693^{3}$ & $0.396^{3}$ &  & 1, 22, 23, 24, 3, 34 \\
        J1416+3444$^\dagger$ & B2~1413+34 & 14:16:04.18 & +34:44:36.39 &  & 81.0$^{*,54}$ & 0.693$^*$ & $0.7^{3}$ & $2.1^{3}$ &  & \phantom{x} \\
        J1434+4236 & B3~1432+428B & 14:34:27.86 & +42:36:20.06 & 0.452$^{25}$ & 68.3$^{*,55}$ & 0.393$^*$ & $<0.074^{*,2}$ & $>1.67^{*,2}$ &  & 2, 22 \\
        J1440+6108$^\dagger$ & VIPS~J14402+6108 & 14:40:17.87 & +61:08:42.88 & 0.445365$^{11}$ & 30.0$^{*,56}$ & 0.171$^*$ & $0.4^{*,1}$ & $0.48^{*,1}$ &  & 22, 23 \\
        J1443+4044 & B3~1441+409 & 14:42:59.32 & +40:44:28.94 &  & 123.4$^{*,57}$ & $^\ddagger$ & $0.292^{3}$ & $1.55^{3}$ &  & 2 \\
        J1508+3423$^\dagger$ & VV~059a & 15:08:05.70 & +34:23:23.00 & 0.045565$^{26}$ & 280.0$^{*,58}$ & 0.247$^*$ & $<0.23^{11}$ & $>0.25^{11}$ &  & \phantom{x} \\
        J1511+0518 & JVAS~J1511+0518 & 15:11:41.27 & +05:18:09.26 & 0.084$^{27}$ & 10.6$^{*,59}$ & 0.017$^*$ & $11.1^{3}$ & $0.778^{3}$ & X$^\mathrm{2,5}$ & 17, 26 \\
        J1559+5924$^\dagger$ & JVAS~J1559+5924 & 15:59:01.70 & +59:24:21.84 & 0.0602$^{24}$ & 11.0$^{*,60}$ & 0.013$^*$ & $<0.15^{*,1}$ & $0.23^{*,1}$ &  & 22, 23 \\
        J1602+5243$^\dagger$ & 4C~+52.37 & 16:02:46.38 & +52:43:58.40 & 0.105689$^{11}$ & 250.0$^{*,61}$ & 0.478$^*$ & $0.15^{11}$ & $1.48^{11}$ & X$^\mathrm{17}$ & \phantom{x} \\
        J1609+2641 & CTD~93 & 16:09:13.32 & +26:41:29.04 & 0.473$^{28}$ & 61.3$^{*,62}$ & 0.362$^*$ & $1.1^{3}$ & $5.44^{3}$ & X$^\mathrm{2,5}$ & 22, 23, 26, 36 \\
        J1645+2536 & PKS~1642+25 & 16:44:59.07 & +25:36:30.64 & 0.588$^{25}$ & 39.0$^{*,63}$ & 0.258$^*$ & $1.0^{*,1}$ & $1.1^{*,1}$ &  & 22, 23 \\
        J1723-6500 & NGC~6328 & 17:23:41.03 & -65:00:36.61 & 0.01443$^{29}$ & 7.0$^{*,64}$ & 0.002$^*$ & $2.7^{3}$ & $4.48^{3}$ & $\gamma^\mathrm{2,3,6,7}$ & 15, 37 \\
        J1734+0926 & PKS~1732+094 & 17:34:58.38 & +09:26:58.26 & 0.735$^{30}$ & 12.8$^{*,65}$ & 0.093$^*$ & $2.3^{5}$ & $1.22^{5}$ &  & 1, 17, 3, 38, 4 \\
        J1735+5049 & CGRaBS~J1735+5049 & 17:35:49.01 & +50:49:11.57 & 0.835$^{31}$ & 8.0$^{*,66}$ & 0.061$^*$ & $6.4^{3}$ & $0.972^{3}$ &  & 31, 39 \\
        J1816+3457 & B2~1814+34 & 18:16:23.90 & +34:57:45.75 & 0.245$^{13}$ & 45.5$^{*,67}$ & 0.174$^*$ & $0.44^{3}$ & $0.983^{3}$ &  & 1, 3 \\
        J1826+1831 & JVAS~J1826+1831 & 18:26:17.71 & +18:31:52.89 &  & 74.0$^{*,68}$ & 0.633$^*$ & $0.308^{8}$ & $1.08^{8}$ &  & 1, 3, 4, 5, 6 \\
        J1826+2708 & B2~1824+27 & 18:26:32.11 & +27:08:07.95 &  & 41.0$^{*,69}$ & 0.351$^*$ & $1.0^{10}$ & $0.34^{10}$ &  & 28, 29 \\
        J1915+6548$^\dagger$ & JVAS~J1915+6548 & 19:15:23.82 & +65:48:46.39 & 0.486$^{32}$ & 36.0$^{*,70}$ & 0.216$^*$ & $0.5^{12}$ & $0.83^{12}$ &  & 5 \\
        J1928+6815 & JVAS~J1928+6814 & 19:28:20.55 & +68:14:59.27 &  & 128.1$^{*,71}$ & $^\ddagger$ & $<0.074^{*,1}$ & $1.04^{*,1}$ &  & 12, 4 \\
        J1939-6342 & PKS~1934-63 & 19:39:25.02 & -63:42:45.62 & 0.183$^{33}$ & 42.6$^{*,72}$ & 0.130$^*$ & $1.4^{3}$ & $15.0^{3}$ & X$^\mathrm{2,5}$ & 4 \\
        J1944+5448$^\dagger$ & S4~1943+54 & 19:44:31.51 & +54:48:07.06 & 0.263$^{34}$ & 48.8$^{*,73}$ & 0.196$^*$ & $0.778^{3}$ & $1.77^{3}$ & X$^\mathrm{2,5}$ & \phantom{x} \\
        J1945+7055 & S5~1946+70 & 19:45:53.52 & +70:55:48.73 & 0.101$^{32}$ & 40.6$^{*,74}$ & 0.075$^*$ & $1.8^{3}$ & $0.929^{3}$ & X$^\mathrm{2,5}$ & 15, 4, 40, 41, 42 \\
        J2022+6136 & S4~2021+61 & 20:22:06.68 & +61:36:58.80 & 0.2266$^{35}$ & 29.0$^{*,75}$ & 0.104$^*$ & $4.086^{3}$ & $2.64^{3}$ & X$^\mathrm{2,5}$ & 4, 43, 44 \\
        J2203+1007 & JVAS~J2203+1007 & 22:03:30.95 & +10:07:42.59 & 1.005$^{36}$ & 11.0$^{*,76}$ & 0.089$^*$ & $4.427^{3}$ & $0.306^{3}$ &  & 1, 17, 3, 39 \\
        J2327+0846$^\dagger$ & NGC~7674 & 23:27:56.70 & +08:46:44.30 & 0.02892$^{37}$ & 1300.0$^{*,77}$ & 0.744$^*$ & $<0.09^{*,2}$ & $>1.0^{*,2}$ & X$^\mathrm{17}$ & \phantom{x} \\
        J2347-1856 & PKS~2344-192 & 23:47:08.63 & -18:56:18.86 &  & 33.4$^{*,78}$ & 0.286$^*$ & $1.8^{4}$ & $0.66^{4}$ &  & 20 \\
        J2355+4950 & TXS~2352+495 & 23:55:09.46 & +49:50:08.34 & 0.23831$^{12}$ & 90.0$^{*,79}$ & 0.337$^*$ & $0.7^{3}$ & $2.93^{3}$ & X$^\mathrm{2,3,4,5}$ & 4, 45, 8, 9 \\
    \enddata
    \tablecomments{
        {
        ($^*$) New measurements (this paper).
        }
        ($^\dagger$) Newly confirmed CSO (this paper).
        ($^\ddagger$) Source may exceed $1\,$kpc linear size if in a certain redshift range; c.f. \cref{sec:noredshift}.
        ($^\mathrm{m}$) Source with multiple phases of activity; c.f. \cref{sec:csos-angular-sizes}.
        {
        Redshifts are taken from the literature; the references are provided in Appendix~\ref{app:csocat}.
        Angular sizes were newly estimated in this work. The frequencies and origins of the maps used for size measurements are listed in Appendix~\ref{app:csocat}. Linear sizes are derived from these angular size measurements.
        Turnover frequencies and flux densities were partially taken from the literature and partially derived from RFC and NED spectra; the references are provided in Appendix~\ref{app:csocat}.
        References for the x-ray and $\gamma$-ray detections are listed in Appendix~\ref{app:csocat}.
        References that discussed some of these sources as CSOs of CSO candidates are listed in Appendix~\ref{app:csocat}.
        }
    }
\end{deluxetable*}
\end{longrotatetable}

%------------------------------------------------------------------------------

\clearpage
%------------------------------------------------------------------------------
%\bibliography{references}
\bibliographystyle{aasjournal}

\end{document}